\documentclass[10pt,showkeys,aps,prl,groupedaddress]{revtex4-2}
\pdfoutput=1

\usepackage[english]{babel}
\usepackage{braket}
\usepackage{bm}
\usepackage[inline]{enumitem}
\usepackage[T1]{fontenc}
\usepackage{graphicx}
\usepackage[ruled,vlined]{algorithm2e}
\usepackage{amsmath}
\usepackage{amssymb}
\usepackage{mathtools}
\usepackage{multirow}
\usepackage{siunitx}
\usepackage[caption=false]{subfig}
\usepackage{pgfplots}

\usepackage{hyperref}
\usepackage{cleveref}

\usepackage{tikz}
\usepackage{tikz-3dplot}
\usetikzlibrary{arrows,automata,calc,positioning,shadows.blur,decorations.pathreplacing,shapes.geometric,decorations.markings,pgfplots.groupplots}
\usetikzlibrary{external}

\definecolor{DARKMAGENTA}{HTML}{AF2F40}
\hypersetup{
    colorlinks=true,
    linkcolor=DARKMAGENTA,
    citecolor=DARKMAGENTA,
    urlcolor=DARKMAGENTA,
}

\crefname{paragraph}{paragraph}{paragraphs}
\Crefname{paragraph}{Paragraph}{Paragraphs}

\usepackage{glossaries}
\newacronym{ace}{ACE}{Atomic Cluster Expansion}
\newacronym{ann}{ANN}{Artificial Neural Network}
\newacronym{krr}{KRR}{Kernel Ridge Regression}
\newacronym{gnn}{GNN}{Graph Neural Network}
\newacronym{mtp}{MTP}{moment tensor potential}
\newacronym{pes}{PES}{potential energy surface}
\newacronym{trs}{TRS}{Translation, Rotation and Scaling}
\glsdisablehyper

\DeclareMathOperator{\Tr}{Tr}

\definecolor{miladblue}{RGB}{174,222,238}
\definecolor{miladpink}{RGB}{230,121,175}

\tikzstyle{class}=[
    rectangle,
    draw=black,
    text centered,
    anchor=north,
    text=black,
    text width=3cm,
    bottom color=white,
    top color=white]

\tikzset{plane/.style n args={3}{insert path={%
#1 -- ++ #2 -- ++ #3 -- ++ ($-1*#2$) -- cycle}},
unit xy plane/.style={plane={#1}{(1,0,0)}{(0,1,0)}},
unit xz plane/.style={plane={#1}{(1,0,0)}{(0,0,1)}},
unit yz plane/.style={plane={#1}{(0,1,0)}{(0,0,1)}},
get projections/.style={insert path={%
let \p1=(1,0,0),\p2=(0,1,0)  in 
[/utils/exec={\pgfmathtruncatemacro{\xproj}{sign(\x1)}\xdef\xproj{\xproj}
\pgfmathtruncatemacro{\yproj}{sign(\x2)}\xdef\yproj{\yproj}
\pgfmathtruncatemacro{\zproj}{sign(cos(\tdplotmaintheta))}\xdef\zproj{\zproj}}]}},
pics/unit cube/.style={code={
\path[get projections];
\draw (0,0,0) -- (1,1,1);
\ifnum\zproj=-1
 \path[3d cube/every face,3d cube/xy face,unit xy plane={(0,0,0)}]; 
\fi
\ifnum\yproj=1
 \path[3d cube/every face,3d cube/yz face,unit yz plane={(1,0,0)}]; 
\else
 \path[3d cube/every face,3d cube/yz face,unit yz plane={(0,0,0)}]; 
\fi
\ifnum\xproj=1
 \path[3d cube/every face,3d cube/xz face,unit xz plane={(0,0,0)}]; 
\else
 \path[3d cube/every face,3d cube/xz face,unit xz plane={(0,1,0)}]; 
\fi
\ifnum\zproj>-1
 \path[3d cube/every face,3d cube/xy face,unit xy plane={(0,0,1)}]; 
\fi
}},
3d cube/.cd,
xy face/.style={fill=blue!10,fill opacity=1.},
xz face/.style={fill=blue!20,fill opacity=1.},
yz face/.style={fill=blue!30,fill opacity=1.},
num cubes x/.estore in=\NumCubesX,
num cubes y/.estore in=\NumCubesY,
num cubes z/.estore in=\NumCubesZ,
num cubes x=1,num cubes y/.initial=1,num cubes z/.initial=1,
cube scale/.initial=0.9,
every face/.style={draw},
/tikz/pics/.cd,
cube array/.style={code={%
 \tikzset{3d cube/.cd,#1}
  \path[get projections];
  \ifnum\yproj=1
   \def\LstX{1,...,\NumCubesX}
  \else 
   \ifnum\NumCubesX>1
    \pgfmathtruncatemacro{\NextToLast}{\NumCubesX-1}
    \def\LstX{\NumCubesX,\NextToLast,...,1}
   \else
    \def\LstX{1}   
   \fi 
  \fi
  \ifnum\xproj=-1
   \def\LstY{1,...,\NumCubesY}
  \else 
   \ifnum\NumCubesY>1
    \pgfmathtruncatemacro{\NextToLast}{\NumCubesX-1}
    \def\LstY{\NumCubesY,\NextToLast,...,1}
   \else
    \def\LstY{1}   
   \fi 
  \fi
  \ifnum\zproj=1
   \def\LstZ{1,...,\NumCubesZ}
  \else 
   \ifnum\NumCubesZ>1
    \pgfmathtruncatemacro{\NextToLast}{\NumCubesX-1}
    \def\LstZ{\NumCubesZ,\NextToLast,...,1}
   \else
    \def\LstZ{1}   
   \fi 
   \def\LstZ{\NumCubesZ,\NextToLast,...,1}
  \fi
  \foreach \X in \LstX
  {\foreach \Y in \LstY
   {\foreach \Z in \LstZ
    {\path (\X-\NumCubesX/2-1,\Y-\NumCubesY/2-1,\Z-\NumCubesY/2-1)
      pic[scale=\pgfkeysvalueof{/tikz/3d cube/cube scale}]{unit cube};}}
  } 
}
}
}

\renewcommand\vec{\bm}

\usepackage{listings}
\DeclarePairedDelimiter\floor{\lfloor}{\rfloor}

\AtBeginDocument{\usepackage{booktabs}}

\begin{document}

\title{Through the eyes of a descriptor: Constructing complete, invertible descriptions of atomic environments}
\author{Martin Uhrin}
\email{martin.uhrin.10@ucl.ac.uk}
\affiliation{Department of Energy Conversion and Storage, Technical University of Denmark, Kgs. Lyngby DK-2800, Denmark}

\date{\today}

\begin{abstract}

In this work we apply methods for describing 3D images to the problem of encoding atomic environments in a way that is invariant to rotations, translations, and permutations of the atoms and, crucially, can be decoded back into the original environment modulo global orientation without the need for training a model.
From the point of view of decoding, the descriptor is optimally complete and can be extended to arbitrary order, allowing for a systematic convergence of the fidelity of the description.
In experiments on molecules ranging from 3 to 29 atoms in size, we demonstrate that positions can be decoded with a 97\% success rate and positions plus species with a 70\% rate of success, rising to 95\% if a second fingerprint is used.
In all cases, consistent recovery is observed for molecules with 17 or fewer atoms.
Additionally, we evaluate the descriptor's performance in predicting the energies and forces of bulk Ni, Cu, Li, Mo, Si and Ge by means of a neural network model trained on DFT data.
When comparing to six machine learning interaction potential methods that use various descriptors and regression schemes our descriptor is found be to competitive, in several cases outperforming well established methods.
The combined ability to both decode and make property predictions from a representation that does not need to be learned lays the foundations for a novel way of building generative models that are tasked with solving the inverse problem of predicting atomic arrangements that are statistically likely to have certain desired properties.

\end{abstract}


\maketitle

{\centering\footnotesize In loving memory of my grandfather, Franti\v{s}ek Karel (1932-2021), whose curiosity for the inner workings of the natural world knew no bounds.\par}

\section{Introduction}
\label{sec:intro}

Predicting the properties of collections of atoms forms a core pillar of many of the technological advances that have become a staple of modern life.
Whether it be drug discovery, materials science, chemical synthesis or chip fabrication, the ability to connect properties to the atomic scale invariably opens the door to accelerated discovery thanks to greater accuracy and transferability when compared to more coarse grained descriptions.
While property prediction is now a mature scientific practice, the inverse problem of predicting atomic arrangements \textit{given} particular target properties is still in its infancy (see \cite{Sanchez-Lengeling2018} for a review of some common methods) and has the potential to have an even greater impact on society than property prediction alone.

Arguably, the most pervasive computational method for the discovery of novel materials and molecules is currently high-throughput screening, which typically involves calculating properties (and often stable atomic arrangements) for many candidates in the hopes of finding some that meet the desired criteria.
Naturally, this often incurs a high computational cost while simultaneously being limited to a predefined set of candidates.
Modern machine learning methods, particularly generative models \citep{Ng2001}, offer the possibility of overcoming some of these limitations by learning patterns within large data sets and proposing novel candidates that are statistically likely to have the desired properties.

Of key importance when building any machine learning model is the representation of the atomic system upon which the learning occurs.
This is reflected in the large volume literature dedicated to the topic (see e.g. \citep{Behler2011,Rupp2012a,Bartok2013a,Behler2011,VonLilienfeld2015,Ferre2015,Shapeev2016,Huo2017,Imbalzano2018,Pozdnyakov2020a,Langer2020,Musil2021}).
What is near universally recognised are the benefits of using representations that possess the same symmetries as the physical laws that govern the atomic interactions.
Specifically, we seek descriptions that possess invariance or equivariance to permutation, translation and rotation.
In addition, for non-linear machine learning methods such as neural networks, it is desirable to have compact descriptors that do not produce redundant information as this greatly improves training efficiency by keeping the dimensionality of the feature space as low as possible.
Finally, for generative models, which rely on inverting latent representations to reproduce atomic structures, it is important that the descriptor be complete such that any set of atomic coordinates maps uniquely onto a point in latent space.
It is this collective set of properties and, perhaps more significantly, the inversion procedure itself that are the focus of this work.

While the topic of symmetry invariant descriptions is now well established in the atomistic modelling community its history goes back significantly further in the fields of image analysis and computer vision.
For these communities, the fundamental objects of interests are pixels, voxels or point clouds, however the methods developed map readily to atomic environments, often by simply placing delta or Gaussian functions on atomic sites.
In this work we borrow and build on methods from 3D image analysis to create a complete descriptor of atomic environments that can be inverted and highlight similarities and connections with existing methods in the atomistic modelling community.
\begin{center}

\begin{tikzpicture}[
  auto, 
  transform shape,
  block/.style={draw, align=center, minimum width=5em, minimum height=3.1em},
  annotation/.style={align=center, font=\footnotesize},
  line join=round,font=\sffamily,3d cube/.cd, num cubes x=1,num cubes y=1,num cubes z=1,every node/.style={scale=5}
  ]

\node at (0,1.2cm) {};
  
\node (env) at (0, 0) {\includegraphics[width=2cm]{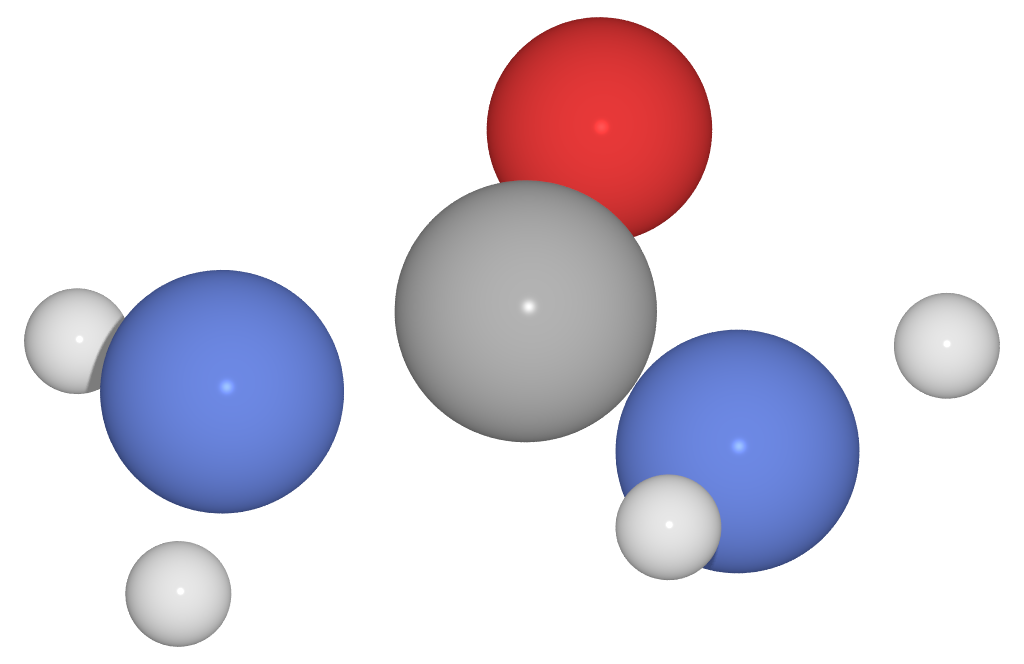}};
\begin{scope}[xshift=3.5cm,scale=1]
 \node[minimum width=2.2cm, minimum height=1.36cm] (features) {};
 \fill[draw=white,outer color=white,inner color=black!60!black,minimum width=3cm] (0.05,0.04) circle (0.15cm);
 \fill[draw=white,outer color=white,inner color=black!60!black,minimum width=3cm] (0.46,-0.24) circle (0.15cm);
 \fill[draw=white,outer color=white,inner color=black!60!black,minimum width=3cm] (0.32,-0.38) circle (0.15cm);
 \fill[draw=white,outer color=white,inner color=black!60!black,minimum width=3cm] (0.88,-0.04) circle (0.15cm);
 \fill[draw=white,outer color=white,inner color=black!60!black,minimum width=3cm] (-0.86,-0.04) circle (0.15cm);
 \fill[draw=white,outer color=white,inner color=black!60!black,minimum width=3cm] (-0.65,-0.51) circle (0.15cm);
 \fill[draw=white,outer color=white,inner color=black!60!black,minimum width=3cm] (-0.56,-0.12) circle (0.15cm);
 \fill[draw=white,outer color=white,inner color=black!60!black,minimum width=3cm] (0.18,0.4) circle (0.15cm);
\end{scope}

\tdplotsetmaincoords{70}{45}
\begin{scope}[xshift=7.6cm, scale=0.42, tdplot_main_coords,local bounding box=moments]
  \path pic{cube array={num cubes x=3,num cubes y=3,num cubes z=3}};
\end{scope}
\tdplotsetmaincoords{85}{190} 
\begin{scope}[xshift=11cm, yshift=-0.7cm, scale=0.42, tdplot_main_coords,local bounding box=invariants]
  \path pic{cube array={num cubes x=1,num cubes z=4}};
\end{scope}

\draw[<->,thick] ($(env.east) + (0.12, 0)$) -> node[annotation,text width=2cm,,yshift=0.1cm] {\Cref{eq:delta-features,eq:gaussian-features}} ($(features.west) + (-0.12, 0)$);


\draw[<-,thick] ($(moments.west) + (-0.19, 0.1)$) -> node[annotation,anchor=south,text width=2cm,yshift=0.1cm] {\Cref{eq:geom_moms,eq:og_moments}} ++(-1.75,0);

\draw[->,thick] ($(moments.east) + (0.28, 0.1)$) -> node[annotation,anchor=south,text width=2cm,yshift=0.1cm] {\Cref{sec:geom_invariants,sec:sph_invariants}} ++(1.9,0);

Inverse links
\draw[->,thick,draw=gray] ($(moments.west) + (-0.28, -0.1)$) -- node[annotation,anchor=north,yshift=-0.1cm] {\Cref{sec:atoms_from_moments}} ++(-1.75, 0);
\draw[<-,thick,draw=gray] ($(moments.east) + (0.19, -0.1)$) -- node[annotation,anchor=north,yshift=-0.1cm] {\Cref{sec:moms_from_invs}} ++(1.9, 0);

\begin{scope}[yshift=-1.4cm]
 \node (envs_label) {Environments};
 \node[anchor=north] at (envs_label.south) {$\{\vec{r}_i, Z_i, \ldots \}$};
 \node (features_label) at (3.5, 0) {Features};
 \node[anchor=north] at (features_label.south) {$f(\vec{r}_i, Z_i, \ldots)$};
 \node (moments_label) at (7.5, 0) {Moments};
 \node[anchor=north] at (moments_label.south) {$c_{nl}^m, m_{stu}, \Omega_{nl}^m$};
 \node (invariants_label) at (11, 0) {Fingerprint};
 \node[anchor=north] at (invariants_label.south) {$\vec{\Phi}$};
\end{scope}

\draw[->,thick] (3.2, -3) -- ++(1,0) node[anchor=west] {$=$ analytic expression};
\draw[->,thick,draw=gray] (3.2, -3.5) -- ++(1,0) node[anchor=west] {$=$ solved numerically};

\end{tikzpicture}
\end{center}

The visual summary above shows a schematic of our method for constructing permutation, translation and rotation invariant fingerprint vectors that can be systematically inverted to recreate the original atomic environment.
References to specific portions of the paper are given to enable the reader to easily skip to the sections of most interest.
Results from numerical experiments can be found in \cref{sec:experiments}.
In order to avoid ambiguity we will use the term \textit{descriptor} to refer to the method and \textit{fingerprint} or \textit{fingerprint vector} (as this evokes the idea of a uniquely identifying artefact) to refer to the object that is produced by a descriptor.

\section{Moment invariants as (local) atomic descriptors}

The moments of a function give a quantitative description of its shape and as such have a long history of use in the world of 2D and 3D image analysis and computer vision.
The first (raw) and second (central) moments of a function are familiar from probability distributions as the mean and variance respectively, while the $n^\text{th}$ moment of a real-valued continuous function $f(x)$ is given by
\begin{equation}
 m_n = \int_{-\infty}^{\infty} x^n f(x) dx\text{.}
 \label{eq:one_d_moments}
\end{equation}
In this context $n$ is typically called the moment's \emph{order} and is equal to the degree of the polynomial onto which $f$ is projected.
More generally, moments (in certain contexts called expansion coefficients) can be defined as the projections of a square-integrable function onto a set of basis functions, $m_{n} = \braket{f | \Psi_n}$.
Many commonly used descriptors of atomic environments are, in fact, functions of moments which produce a fingerprint vector, denoted here as $\vec{\Phi}$.

For the purposes of faithfully encoding and, crucially, reconstructing atomic environments we are particularly interested in descriptors that have the following properties:
\begin{enumerate}
 \item Invariance to global translations and rotations, and permutations of atom labels.
 \item Orthonormality of the basis functions i.e. $\braket{\Psi_i | \Psi_j} = \delta_{ij}$ \footnote{It is not strictly necessary to work with an orthogonal basis but in this case the basis functions need be linearly independent and invertible for direct synthesis to be possible.}.  Such basis functions lead to an optimal reconstruction of $f$ in the sense that the contribution to the mean square error decreases with each successive moment order $n$ \footnote{Note that for particular classes of image function there may still be differences in convergence rates for different choices of orthonormal basis functions.}.
 \item Completeness of the basis functions.  If, for any function $f \in L^2$, the following condition is satisfied
 \begin{equation}
  \lim_{n_\text{max} \to \infty} \left\Vert f - \sum_{n=0}^{n_\text{max}} m_n \Psi_n \right\Vert^2 = 0 \text{,}
  \label{eq:completeness}
 \end{equation}
 we say that the basis, $\Psi_n$, is complete as $f$ can be approximated arbitrarily closely.
  \item A smooth, injective, invariants function, $\vec{m}  \mapsto \vec{\Phi}(\vec{m})$, that maps any two vectors of moments ($\vec{m} = (m_1, \ldots, m_{n_\text{max}})$) \emph{not} related by a global rotation onto distinct fingerprint vectors.
\end{enumerate}

In what follows the ultimate goal is to identify procedures for generating systematically improvable descriptions that satisfy, as much as possible, the aforementioned criteria and that can be used to both predict properties \emph{and} to recreate atomic environments from a finite set of invariants.
We call such descriptors moment invariants (local) atomic descriptors (\textsc{MILAD}).

\subsection{Geometric moments}
\label{sec:geom_invariants}

The extension of \cref{eq:one_d_moments} to three dimensions is trivial; using three indices, $s$, $t$ and, $u$, the moments of a three-dimensional function, $f(x_1, x_2, x_3)$, are given by
\begin{equation}
 m_{stu} = \int_{-\infty}^\infty \int_{-\infty}^\infty \int_{-\infty}^\infty x_1^s x_2^t x_3^u f(x_1, x_2, x_3) dx_1 dx_2 dx_3 \text{,}
 \label{eq:geom_moms}
 \end{equation}
where the order is now $s + t + u$.
These are the, so-called, geometric moments and are widely used in 3D image analysis where the function $f(\vec{x})$, with $\vec{x} = (x_1, x_2, x_3)^T$, is typically a set of discrete voxels of varying intensity \citep{Flusser2016}.
Importantly, one can prove \citep{Reiss1993} that for piecewise continuous $f(\vec{x})$ with compact support, all moment orders exist and that $f(\vec{x})$ is uniquely determined by the set of all moments to infinite order which are themselves uniquely determined by $f(\vec{x})$.  
While this is formally true only in the limit of infinite order, we will show later that it is possible to use a related set of orthogonal moments to recover atomic environments from finitely many moments and, indeed, rotation invariants thereof.

In order to use moments to represent atomic configurations, we must make a choice for the form of $f(\vec{x})$ and a convenient one is that of a sum (thereby inducing permutational invariance) of feature functions centred on each atom, perhaps the simplest of which is a set of $N$ atom centred delta functions with weights $w_i$
\begin{equation}
 f(\vec{x}) = \sum_i^N w_i \delta (\vec{x} - \vec{r}_{i}) \text{,}
 \label{eq:delta-features}
\end{equation}
where $\vec{r}_{i}$ is the position of the $i^\text{th}$ atom and the weights can either all be equal to one or be used to encode additional information, for example the atomic species.
This choice of $f(\vec{x})$ is particularly appealing as the moments can be trivially computed as
\begin{equation}
 m_{stu} = \sum_i^N w_i \vec{r}_{i,1}^s \vec{r}_{i,2}^t \vec{r}_{i,3}^u \text{.}
 \label{eq:delta-moments}
\end{equation}
These moments, $m_{stu}$, can be grouped by order into tensors from which rotation invariants can be easily derived.
This forms the basis of much early work in image analysis and more recently in the atomistic modelling community, most notably in the work of \citet{Shapeev2016} and \citet{Zaverkin2020}.


Another commonly used feature function (see e.g. \citep{Sommer2007,Bartok2013a,Willatt2019}) is that of a three-dimensional Gaussian, in which case $f(\vec{x})$ becomes
\begin{equation}
 f(\vec{x}) = \sum_i^N{ \frac{w_i}{\sigma_i^3 \sqrt{(2\pi)^3}} \exp \left(  -\frac{1}{2} \left( \frac{\vec{x} - \vec{r}_{i}}{\sigma_i} \right)^2 \right) } \text{,}
 \label{eq:gaussian-features}
\end{equation}
where both $w_i$ and the Gaussian widths, $\sigma_i$, can be used to encode information about atom $i$ \footnote{In fact, $\sigma$ can even be a covariance matrix, giving the atomic feature an asymmetric shape.  Something that can be exploited to encode anisotropic properties such as spins or chemical shifts}.
The analytic integrals $\iiint x_1^s x_2^t x_3^u f(x_1, x_2, x_3) dx_1 dx_2 dx_3$, and therefore the moments, can be found relatively easily \footnote{The first few orders are given in the Supplemental Material \cite{SI}, the MILAD source code contains explicit forms for up to order 16 inclusive.} and are fast to compute.

The range of possible forms for $f(\vec{x})$ is infinite and the choice is somewhat arbitrary, and can therefore be guided by seeking forms that localise the atomic positions well and that have integrals that can be efficiently evaluated while providing enough degrees of freedom to encode the desired properties, e.g. atomic species, spins, chemical shieldings, or anything else relevant to the property being predicted.

\subsubsection*{Rotation invariants}

Thus far, $m_{stu}$ are permutationally invariant due to choosing $f(\vec{x})$ that are sums of feature functions.  
Translational invariance can be achieved by normalisation, that is by choosing a particular reference point to be the origin of our coordinate system.  
In the case of local atomic descriptors, this is typically taken to be the position of each atom in turn, around which a cutoff sphere is projected and all the atoms that lie in its interior are included in $f_i (\vec{x})$.  
This leads to a set of fingerprints, one for each atom, which can be used directly as inputs to a suitable fitting algorithm to predict local properties such as forces or chemical shifts \citep{Paruzzo2018}.
Alternatively, the set of local fingerprints can be used together to predict global properties such as total energy or band gap, either by combining them to form a new descriptor or directly (e.g. as is done in Behler-Parinello neural network potentials \cite{Behler2011}).
For describing global environments such as entire molecules, nanoribbons, clusters, etc., the centre of mass or average position can be used as the reference point or it can be chosen more specifically to fit the problem at hand.

The remaining invariance to tackle is that of rotation.
The moments, $m_{stu}$, are not, themselves, rotationally invariant, however various methods have been developed all yielding polynomials of moments that are.
Some of the earliest work on rotation invariants from moments in three dimensions was carried out by \citet{Sadjadi1980} however it was in the work of \citet{Suk2011} that a general solution was first presented by extending the ideas of \citet{Cyganski1986} on 2D moment tensors to 3D.

Using Einstein notation, we start with the definition of the moment tensor
\begin{equation}
 M^{i_1 i_2 \dots i_k} = \int_{-\infty}^{\infty} \int_{-\infty}^{\infty} \int_{-\infty}^{\infty} x^{i_1} x^{i_2} \dots x^{i_k} f(x^1, x^2, x^3) dx^1 dx^2 dx^3\text{,}
\end{equation}
which has $k$ indices and each $i_j$ can be either 1, 2, or 3 corresponding to the $x_1$, $x_2$ or $x_3$ coordinate respectively.  
The total number of repetitions of each corresponds to the exponents $s$, $t$ and $u$ in \cref{eq:geom_moms}, giving the relation between $M^{i_1 i_2 \dots i_k}$ and the geometric moments $m_{stu}$.
For the case of rotations (as opposed to general affine transformations, which can also be symmetrised \cite{Kostkova2021} but are more complicated and less relevant for atomic systems) the moment tensor can be treated as a Cartesian tensor, in which case the distinction between covariance and contravariance is lifted and $M_{i_1 i_2 \dots i_k}$ transforms simply as
\begin{equation}
 \hat{M}_{\alpha_1,\alpha_2,\dots,\alpha_k} = Q_{\alpha_1 i_1} Q_{\alpha_2 i_2} \dots Q_{\alpha_k i_k} M_{i_1,i_2,\dots,i_k}\text{,}
\end{equation}
for a given orthonormal rotation matrix $Q_{ij}$.  It can be shown that the total contraction of a Cartesian tensor, e.g. $M_{kk}$, and total contractions of products of Cartesian tensors, e.g. $M_{kl}M_{kl}$, $M_{kl}M_{lm}M_{mk}$, etc., are all rotation invariants \citep{Flusser2016}.  
The first two Suk-Flusser invariants \footnote{Note that typically the Suk-Flusser invariants are given in terms of the central geometric moments and scale normalised by dividing by an appropriate power of the $m_{000}$ moment which encodes the total `mass' of the system.  For atomic systems, however, scale invariance is rarely desired.} expressed in terms of geometric moments are:
\begin{align*}
 \Phi_1 = & M_{kk} = m_{200} + m_{020} + m_{002} \\
  \Phi_2 = & M_{kl}M_{kl} = m_{200}^2 + m_{020}^2 + m_{002}^2 + 2 m_{110}^2 + 2 m_{101}^2 + 2 m_{011}^2 \text{.}
\end{align*}
This result can be understood intuitively by considering the case of atom centred delta functions (\cref{eq:delta-features}).
Using \cref{eq:delta-moments} we have that
\begin{equation}
 \Phi_1 = \sum_{i = 1}^N w_i(\vec{r}_{i,1}^2 + \vec{r}_{i,2}^2 + \vec{r}_{i,3}^2 ) = \sum_{i = 1}^N w_i \vec{r}_{i} \cdot \vec{r}_{i}\text{,}
\end{equation}
which is nothing more than the sum of the weighted dot products of each atom's position vector with itself, a quantity that is plainly rotationally invariant.
We may also recognise this as the trace of the inertia tensor of a rigid body composed of point masses, $\Tr(\mathbf{I})$, a quantity that is well known in engineering applications as the \textit{first principle invariant}.
For the case of $\Phi_2$ we have
\begin{align*}
 \Phi_2 = & \left( \sum_{i=1}^N w_i \vec{r}_{i,1}^2 \right)^2 + \left( \sum_{i =1}^N w_i \vec{r}_{i,2}^2 \right)^2 + \left( \sum_{i = 1}^N w_i \vec{r}_{i,3}^2 \right)^2 +\\
  & 2 \left( \sum_{i = 1}^N w_i \vec{r}_{i,1} \vec{r}_{i,2} \right)^2 + 2 \left( \sum_{i = 1}^N w_i \vec{r}_{i,1} \vec{r}_{i,3} \right)^2 + 2 \left( \sum_{i = 1}^N w_i \vec{r}_{i,2} \vec{r}_{i,3} \right)^2\text{,}
\end{align*}
which is the \textit{second main invariant}, $\Tr(\mathbf{I}^2)$.

This procedure can be continued in a similar fashion to yield as many invariants as are needed to faithfully describe the atomic environments under investigation, however, care must be taken to avoid redundant invariants which can take several forms as discussed later.
The procedure outlined thus far has much in common with that used to build invariants for \glspl{mtp} \cite{Shapeev2016,Gubaev2018} and Gaussian moment invariants \cite{Zaverkin2020}, differing primarily in leaving the choice of $f(\vec{x})$ free.

\subsection{Spherical harmonic based moments}

While geometric moments are fast to calculate, their numerical properties make them less suitable for some applications, and other choices of basis function can yield better performance, particularly when it comes to reconstruction.
Another commonly used set of functions are the spherical harmonics, which yield complex moments defined as
\begin{equation}
 c_{nl}^m = \int_0^{2\pi} \int_0^\pi \int_0^\infty R_{nl}(r) \overline{Y_l^m (\theta, \varphi)} f(r, \theta, \varphi) r^2 \sin{\theta} dr d\theta d\varphi \text{,}
\end{equation}
where $R_{nl}(r)$ are a set of real-valued radial basis functions and
 $r^2 = x_1^2 + x_2^2 + x_3^2$, $\theta = \arccos(x_3 / r)$ and $\varphi = \arctan(x_2 / x_1)$.
Here we use the definition
\begin{equation}
 Y_l^m(\theta, \varphi) = \sqrt{\frac{(2 l + 1)}{4 \pi}\frac{(l - m)!}{(l + 1)!}} P_l^m(\cos(\theta)) e^{i m \varphi},
\end{equation}
where $P_l^m$ are the \textit{associated Legendre functions}.

Spherical harmonics are particularly well suited to deriving rotation invariants and, as such, have seen use 3D image analysis for some time.
Pioneering work was carried out by \citet{Lo1989} who used group-theoretical methods to find rotation invariants for 3D object identification and positioning.  
A similar approach was followed by \citet{Canterakis1999} and later \citet{Novotni2004} who both used Zernike polynomials \citep{Zernike1934} as radial basis functions to arrive at 3D rotation invariants.
More recent work by \citet{Suk2015} has demonstrated a systematic ways of arriving at independent invariants from complex moments which we detail below.

The atomistic modelling, bioinformatics and computational chemistry communities have also adopted such rotation invariants both for predicting properties and, in particular, learning \glspl{pes}.
Some of the earliest examples can be found in work on molecules \citep{Max1988,Duncan1993,Duncan1993b,Duncan1995,Morris2005,Ritchie1999,Cai2002,Morris2005,Sommer2007,Grandison2009} while the last decade has seen widespread adoption of spherical harmonic based descriptors, often coupled with advanced machine learning techniques, to predict properties of solids as well.
The well known SOAP descriptor \citep{Bartok2013a} was one of the first and shares much in common with the invariants of \citet{Canterakis1999}.
Recent work in the atomistic modelling community has also turned to the use of polynomials of moments very similar to those described here, showing a growing convergence between the directions taken by the 3D image analysis and atomistic modelling communities \citep{Drautz2019,Bachmayr2019,Drautz2020,Grisafi2020,Nigam2020}.

\subsubsection{Radial basis}

Perhaps the simplest radial basis is $R_n(r) = r^n$ which gives rise to a set of moments which are widely used in 2D and 3D image analysis (see e.g. \citep{Abu-Mostafa1985,Bhattacharya1997,Flusser2006,Flusser2009}).
Using spherical harmonics in Cartesian form the, so called, \textit{3D complex moments} are given by
\begin{equation}
 c_{nl}^m = \int_{-\infty}^{\infty} \int_{-\infty}^{\infty} \int_{-\infty}^{\infty} r^n Y_l^m(x_1, x_2, x_3) f(x_1, x_2, x_3) dx_1 dx_2 dx_3 \text{.}
\end{equation}
The first few spherical harmonics are
\begin{align*}
 Y_0^0(x_1, x_2, x_3) &= \frac{1}{2} \sqrt{\frac{1}{\pi}},\\
 Y_1^0(x_1, x_2, x_3) &= \frac{1}{2} \sqrt{\frac{3}{\pi}} \cdot \frac{x_3}{r}, \\
 Y_1^1(x_1, x_2, x_3) &= -\frac{1}{2} \sqrt{\frac{3}{2\pi}} \cdot \frac{(x_1 + ix_2)}{r}, \\
 Y_2^0(x_1, x_2, x_3) &= \frac{1}{4} \sqrt{\frac{5}{\pi }} \cdot \frac{(3 x_3^2 - r^2)}{r^{2}},\\
  Y_2^1(x_1, x_2, x_3) &= -\frac{1}{2} \sqrt{\frac{15}{2\pi }} \cdot \frac{( x_1+ ix_2) x_3}{r^{2}},\\
 Y_2^2(x_1, x_2, x_3) &= \frac{1}{4} \sqrt{\frac{15}{2\pi }} \cdot \frac{(x_1+ix_2)^{2}}{r^{2}},
\end{align*}
where $Y_l^{-m} = (-1)^m \overline{Y_l^m}$.
When the basis consists of polynomials in $(x_1, x_2, x_3)$ the moments provide a complete and independent description of $f$, we therefore make the restriction that $n - l$ be even and $l \le n$, such as to cancel the factor of $r^{-l}$ that comes from the spherical harmonics.
This basis, is thus, a set of complex homogenous polynomials of degree $n$ in $(x_1, x_2, x_3)$.

While 3D complex moments are fast to calculate, they have some notable disadvantages compared to orthogonal moments.
For one, they posses high dynamic range which can lead to greater numerical errors, in addition indirect methods must to be used to perform reconstruction (see e.g. \citep{Ghorbel2006}).

Orthogonal radial functions have a region of orthogonality which we take to be $0 \le r \le 1$, this requires that atomic environments be scaled appropriately to fit within the unit sphere.  
In general, the corresponding \textit{orthogonal moments} are given by
\begin{equation}
 c_{nl}^m = \eta_{nlm} \int_0^{2\pi} \int_0^\pi \int_0^1 R_{nl}(r)\overline{Y_l^m (\theta, \varphi)} f(r, \theta, \varphi) r^2 sin \theta dr d\theta d\varphi \text{,}
 \label{eq:og_moments}
\end{equation}
where $\eta_{nlm}$ is an normalisation constant.
Orthogonal basis functions make it trivial to reconstruct an approximation to the original function given a finite set of moments.
The reconstruction up to order $n_\text{max}$ is given by
\begin{equation}
 \tilde{f}(r, \theta, \varphi) = \sum_n^{n_\text{max}} \sum_{l=0}^n \sum_{m=-l}^l c_{nl}^m R_{nl}(r) Y_l^m (\theta, \phi)\text{.}
 \label{eq:reconstruction}
\end{equation}
This expansion minimises the mean-square error to the original function and can be systematically converged making it ideally suited for the task of reconstruction.

\subsubsection{Rotation invariants}
\label{sec:sph_invariants}

Here we give a brief description of how to arrive at a set of rotation invariants to arbitrary correlation order, consisting of polynomials of complex moments which can subsequently be reduced to an independent set of rotation invariants.  
The flexibility to expand to arbitrary order is particularly appealing given the recent evidence that three and four-body atom-centred features are insufficient to unambiguously describe all possible atomic environments \cite{Pozdnyakov2020a}.
We follow the procedure of \citet{Suk2015} who use ideas from \citet{Lo1989} and encourage the reader to refer to these sources for a more complete description.

Spherical harmonics transform as
\begin{equation}
 Y_l^m(\mathbf{Q}^{-1} \vec{x}) = \sum_{m' = -l}^l D_{m'm}^{l}(\mathbf{Q}) Y_l^{m'}(\vec{x})\text{,}
 \label{eq:yml_transform}
\end{equation}
when rotated by an arbitrary rotation matrix $\mathbf{Q}$ where $D_{m'm}^{l}$ are the, so-called, Wigner D-functions.
It is common to express the spherical harmonics as a vector, $\vec{Y}_l(\vec{x}) = (Y_l^{-l}(\vec{x}), Y_l^{-l + 1}(\vec{x}), \cdots, Y_l^{l}(\vec{x}))^T$ and $D_{m'm}^l$ as $(2l + 1) \times (2l + 1)$ dimensional unitary matrices, called Wigner D-matrices, that form an irreducible representations of the group of three-dimensional rotations, SO(3).
The elements of $\mathbf{D}^l$ are given by
\begin{equation}
 D_{mm'}^l = \braket{Y_l^m|\hat{R}(\alpha \beta \gamma)|Y_l^{m'}}\text{,}
\end{equation}
where $\hat{R}(\alpha \beta \gamma)$ is the rotation operator parameterised by three Euler angles.  

For a given $l$, the corresponding spherical harmonics $\vec{Y}_l$ form a basis of the irreducible representation $\mathbf{D}^{l}$.
Invariants to rotation can be found by identifying one-dimensional irreducible subspaces that transform according to $\mathbf{D}^0$.
$Y_0^0$ is such an invariant, however spherical harmonics corresponding to irreducible representations with $l \ne 0$ are not.
To find further invariants we can, for example, combine basis functions corresponding to representations $\mathbf{D}^{j_1}$ and $\mathbf{D}^{j_2}$ by taking their tensor product.  Maschke's theorem tells us that the resulting representation can be expressed as the direct sum of irreducible representations
\begin{equation}
 \mathbf{D}^{j_1} \otimes \mathbf{D}^{j_2} = \bigoplus_{l=|j_1 - j_2|}^{j_1 + j_2} \mathbf{D}^{l}\text{,}
\end{equation}
where the right hand side is a block diagonal matrix and the upper-left submatrix, $\mathbf{D}^0$, corresponds to a scalar that is a rotation invariant.
In general, the connection between basis functions $\varphi_{j_1}^i$ and $\varphi_{j_2}^i$ corresponding to the irreducible representations $\mathbf{D}^{j_1}$ and $\mathbf{D}^{j_2}$ and the basis functions, $\Psi_j^k$, of a particular $\mathbf{D}^l$ is
\begin{equation}
    \Psi_l^k = \sum_{m = \max{(-j_1, k - j_2)}}^{\min{(j_1, k + j_2)}} \braket{j_1, j_2, m, k - m|l, k} \varphi_{j_1}^m \varphi_{j_2}^{k-m}
    \label{eq:irreducible_basis}
\end{equation}
where $\braket{j_1, j_2, m, k - m|l, k}$ are Clebsch-Gordan coefficients.
As the radial basis functions do not couple to $m$, the tensor products of moment vectors, $\vec{c}_{nl} = (c_{nl}^{-l}, c_{nl}^{-l + 1}, \cdots, c_{nl}^{l})^T$, are also related by \cref{eq:irreducible_basis}.
This led \citeauthor{Lo1989} to define \textit{composite complex moment forms}
\begin{equation}
 c_n(l_1, l_2)_l^k = \sum_{m = \max{(-l_1, k - l_2)}}^{\min{(l_1, k + l_2)}} \braket{l_1, l_2, m, k - m|l, k} c_{nl_1}^m c_{nl_2}^{k-m}\text{,}
 \label{eq:lo-don}
\end{equation}
which can be used as a mathematical tool to calculate moments in the basis of representation $\mathbf{D}^l$.
Modern equivariant neural networks \citep{Thomas2018,Kondor2018c,Fuchs2020,Miller2020,Geiger2021} use similar ideas to ensure that vectorial or tensorial inputs are propagated through the network in a symmetry preserving way.

In the atomistic modelling community it is common to speak in terms of the \textit{correlation order} of an invariant, labelled $\nu$, which is given by the number of terms involved in a tensor products, while the signal processing community typically uses the term \textit{degree} (or order) in reference the polynomial degree in moments.
The two are equivalent and here will use the former.
Using \cref{eq:lo-don} a set of rotation invariants can be defined as follows.

\paragraph*{$\nu = 1$ invariants}

$c_{n0}^0$ are invariants corresponding to $\mathbf{D}^0$, where $n$ is even.

\paragraph*{$\nu = 2$ invariants}

$c_n(l, l)_0^0$ are invariants corresponding to $(\mathbf{D}^{l} \otimes \mathbf{D}^{l})^0$, given by
\begin{equation}
\sum_{m = -l}^{l} \braket{l, l, m, - m|0, 0} c_{nl}^m c_{nl}^{-m} = \frac{1}{\sqrt{2l + 1}} \sum_{m=-l}^l (-1)^{l - m} c_{nl}^m c_{nl}^{-m}\text{,}
\end{equation}
where $n - l$ must be even.  This is often called the power spectrum and is used by a number of existing descriptors.

\paragraph*{$\nu = 3$ invariants}

Further invariants can be arrived at by combining the bases of more representations, for example, $((\mathbf{D}^{l_1} \otimes \mathbf{D}^{l_2})^l \otimes \mathbf{D}^l)^0$ invariants are given by
\begin{equation}
c_{n_1}(l_1,l_2)_l c_{n_2} = \frac{1}{\sqrt{2l + 1}} \sum_{k=-l}^l (-1)^{l - k} c_{n_1}(l_1, l_2)_l^k c_{n_2 l}^{-k}\text{,}
\end{equation}
where $n_1 - l_1$, $n_1 - l_2$ and $n_2 - l$ are even. Swapping $l_1$ and $l_2$ produces the same invariant and therefore we take $l_2 \leq l_1$.  This is often called the bispectrum.

\paragraph*{$\nu  = 4$ invariants}

Continuing in this fashion $((\mathbf{D}^{l_1} \otimes \mathbf{D}^{l_2})^l \otimes (\mathbf{D}^{l_3} \otimes \mathbf{D}^{l_4})^l)^0$ gives the invariants
\begin{equation}
c_{n_1}(l_1,l_2)_l c_{n_2}(l_3,l_4)_l = \frac{1}{\sqrt{2l + 1}} \sum_{k=-l}^l (-1)^{l - k} c_{n_1}(l_1, l_2)_l^k c_{n_2}(l_3, l_4)_l^{-k}\text{,}
\end{equation}
where $n_1 - l_1$, $n_1 - l_2$, $n_2 - l_3$, and $n_2 - l_4$ are even and to avoid duplicates we take $l_1 \leq n_1$, $l_1 - l \leq l_2 \leq l_1$, $l_3 \leq n_2$, and $l_3 - l \leq l_4 \leq l_3$.

This procedure may be extended to yield as many invariants as are needed however there will invariably be many that are dependent or identical to others which must then be reduced to an independent set as outlined below.
The procedure of generating successively higher correlation order invariants is also at the core of the \gls{ace} \citep{Drautz2019,Drautz2020} scheme which uses linear regression to expand the energy of a collection of atoms in a basis of such invariants.
Recent results \citep{Lysogorskiy2021a,Zeni2021} show very good performance both in terms of accuracy and computational efficiency.

\subsection{Independent invariants}
\label{sec:reducing-invariants}

Be they based on geometric or complex moments, a set of rotation invariants can be arrived at by na\"{i}vely carrying out one of the procedures outlined above, however this will invariably lead to a set that is overcomplete.
Certain types of dependency can be eliminated during generation, however, others have to be checked \textit{post facto} which can have a high algorithmic complexity.
\citet{Suk2011} provide a useful analysis which begins with the following classification of redundancies:
\begin{enumerate*}
    \item \textit{zero invariants},
    \item \textit{identical invariants},
    \item \textit{direct products} (where an invariant is a product of previously found invariants),
    \item \textit{linear combinations}, and
    \item \textit{polynomial dependencies.}
\end{enumerate*}
In general, the first three can be eliminated relatively easily by brute force using a suitable symbolic mathematics library.
Some practical methods for eliminating linear and polynomial dependencies are described below.

\subsubsection{Linear dependencies}

In general, linear combinations can be detected by calculating the column rank of the matrix representing the system of equations given by the invariants at each correlation order, where the columns consist of the moment coefficients of each invariant.
If the number of columns exceeds the rank, then the number of invariants can be reduced by performing a singular value decomposition and discarding those with zero singular values.
If a resulting matrix can be found that has full rank, then the invariants are linearly independent.
While this procedure works, in principle, for any set of invariants it can suffer from problems of numerical instability, specifically when the polynomial coefficients are not integers.

An alternative approach for eliminating linear dependencies amongst invariants from geometric moments is to identify unique generating graphs from the networks corresponding to the tensor contractions \citep{Suk2011}.
The approach of using tensor networks to construct invariants is also used by \citet{Zaverkin2020} for their Gaussian moments descriptor.

For invariants from spherical harmonics many linear dependencies can be eliminated by using standard results from the coupling of angular momenta.
For example, as in \cref{sec:sph_invariants}, by restricting the $l$ indices involved in a tensor product to be in sorted order e.g. $l_1 \ge l_2 \ge \cdots \ge l_\text{max}$, a large number of dependencies are avoided.
Furthermore, if any two or more two irreps involved in a tensor product share the same value of $l$ then the $n$ indices may also be similarly sorted to avoid further dependencies.
This scheme is employed as part of the construction of NICE \citet{Nigam2020} descriptors.
Finally, the work of \citet{Bachmayr2019}, which underpins the \gls{ace} descriptor, gives a numerically stable algorithm for generating a fully linearly independent set of invariants to arbitrary correlation order using a complete analysis of the permutational symmetries amongst invariants.

\subsubsection{Polynomial dependencies}
\label{sec:poly_deps}

A set of invariants with no polynomial dependencies is called independent or algebraically complete.
General procedures to generate such a set are prohibitively expensive, even for relatively low correlation orders \citep{Guo2018a}.
Nevertheless, a number of algorithms that try to reduce some or all such dependencies have been proposed \cite{Langbein2009,Hickman2012,Kostkova2021}, and while they are not guaranteed to produce a certifiably independent set, in practice it is sufficient to test their algebraic independence on a representative set of moments.
Not all methods will produce the same independent invariants due to arbitrary choices that are made, but in all cases the total number of independent invariants to a given order should equal the number of moments minus the number of degrees of freedom that are being symmetrised.

Despite the algorithmic complexity this step is important for non-linear regression schemes as the reduction in the number of invariants can be significant, greatly reducing the dimensionality of the feature space.
\Cref{tab:num_invariants} shows the number of invariants generated by the procedure in \cref{sec:sph_invariants} up to $l_\text{max} = 7$ and the corresponding number of linearly and algebraically independent invariants.
\begin{table}
\begin{tabular}{l r r r r r r}
 \hline\hline
 $l_\text{max}$ & 2 & 3 & 4 & 5 & 6 & 7 \\
 \hline
 All & 4 & 16 & 49 & 123 & 280 & 573 \\
 Linearly indep. & 3 & 13 & 37 & 100 & 228 & 486 \\
 Algebraically indep. & 3 & 13 & 28 & 49 & 77 & 113 \\
 \hline\hline
\end{tabular}
\caption{
The number of invariants to translation and rotation from complex moments with the corresponding number of linearly and algebraically invariants up to a given $l_\text{max}$ (here the conditions that $l \le n$ and $n - l$ is even are assumed).}
\label{tab:num_invariants}
\end{table}

In this work we use invariants from complex moments made independent by the method of \citet{Kostkova2021} which is based on earlier work by \citet{Langbein2009,Hickman2012}, and can be summarised as follows.
Given a set of $n_k$ invariants $\{I_1, \ldots, I_{n_k}\}$, if there are dependencies then we can write any invariant as some function of the others, e.g. 
\begin{equation}
I_1(c_{nl}^m) = f(I_2(c_{nl}^m), \ldots, I_{n_k}(c_{nl}^m)).
\label{eq:dep_invs}
\end{equation}
Taking the derivative with respect to a moment, we get
\begin{equation}
 \frac{\partial I_1 (c_{nl}^m)}{\partial c_{nl}^m} = \frac{d f(I_2 (c_{nl}^m), \ldots, I_{n_k} (c_{nl}^m))}{d c_{nl}^m} = \sum_{\beta = 2}^{n_k} \frac{\partial f}{\partial I_\beta} \frac{\partial I_\beta}{\partial c_{nl}^m}\text{,}
\end{equation}
which shows that the derivatives of the dependent invariants are, themselves, linearly dependent.
This gives an intuitive way to understand the Jacobian criterion for algebraic independence which states that if the invariants are dependent then, by rearranging \cref{eq:dep_invs}, there must exist a function, $F$, such that
\begin{equation}
 F(I_1, \ldots, I_{n_k}) = 0\text{.}
\end{equation}
For convenience we sort the moments, $c_{nl}^m$, in lexicographic order such that we can use a single index, $\alpha = (n, l, m)$, which runs from 1 to $n_p$.
Using this notation, the derivatives of $F$ are
\begin{equation}
 \frac{\partial F(I_1, \ldots, I_{n_k})}{\partial c_\alpha} = \sum_{\beta = 1}^{n_k} \frac{\partial F(I_1, \ldots, I_{n_k})}{\partial I_\beta} \frac{\partial I_\beta}{\partial c_\alpha} = 0,
 \label{eq:jacobi}
\end{equation}
where we have a known, $n_k \times n_p$, Jacobian matrix $A_{\alpha\beta} = \frac{\partial I_\beta}{\partial c_\alpha}$ and an unknown coefficients vector $b_\beta = \frac{\partial F(I_1, \ldots, I_{n_k})}{\partial I_\beta}$ of length $n_k$.
If all of the invariants are independent then the only solution is $b_\beta = 0\text{, }\forall \beta$.
Accordingly, if the Jacobian has full column rank then the invariants are algebraically independent.
Conversely, if the rank $n_r$ is less than $n_k$ then there are only $n_r$ independent invariants.

As the column rank of $A_{\alpha\beta}$ cannot be determined analytically this step must be performed numerically for several generic sets of moments and the maximum rank found is taken to be $n_r$.
If $n_r$ is less than $n_k$ then some criterion must be used to choose which invariants to keep.
\citeauthor{Kostkova2021} follow the following procedure:
\begin{enumerate}
 \item Order the invariants first by correlation order and then by number of terms (this guarantees that any subset of invariants up to arbitrary $l_\text{max}$ is also algebraically complete)
 \item Add one invariant at a time to the set of invariants.
 \item Check the new rank of $A_{\alpha\beta}$ by testing against the representative set of moments (in their case 5 were used) and if it increases by one, keep the invariant, otherwise discard it.
\end{enumerate}
While this procedure is relatively computationally expensive, it need only be carried out once.
It is important to either validate the tolerance of the method used to determine the rank or to use a library that supports exact arithmetic to ensure that the outcome is not affected by numerical issues.

The above procedure works for any set of invariants (irrespective of the basis functions used), while for  spherical harmonic based invariants the procedure of \citet{Nigam2020} offers an alternative that does not rely on numerical evaluation and therefore avoids the associated issues of choosing a tolerance and having to use a representative set of moments, however, their methods will not necessarily identify all polynomial dependencies.

\subsubsection{Data driven invariants reduction}

The invariants reduction analysis thus far focused solely on eliminating redundancies that do not lead to any information loss over all possible signals (i.e. sets of moments).
There are, however, data driven methods for reducing the number of invariants that are based on features found in a particular data set and allow the descriptor to retain discriminative power amongst the entries of that set, potentially sacrificing universality.
It may even be possible to use such techniques to find a reduced set of invariants that work well for all signals that are sums of atomic feature functions (with a corresponding loss of discriminative power for other square-integrable functions).
As this work is focused on universal descriptors we will not detail any particular approach here but we refer the reader to \cite{Nigam2020,Goscinski2021a} for two examples of such techniques.

\subsection{This work}

Our codebase contains support for delta and Gaussian feature functions as well as a set of 1185 linearly independent and 962 algebraically independent invariants from geometric moments up to 16$^\text{th}$ order.
The code is fully modular and can be easily extended with new invariants, basis functions or feature functions.

\subsubsection*{Zernike moments}

For invariants based on spherical harmonics we use Zernike functions as the radial basis, however other choices have been proposed including Bessel \citep{Kocer2020} and Chebyshev \citep{Bachmayr2019} functions which may have different numerical characteristics.
(See \citet{Goscinski2021a} for an excellent review of the performance of various radial basis functions.)
The 3D Zernike polynomials are defined as
\begin{equation}
 Z_{nl}^m(r, \theta, \varphi) = R_{nl}(r) Y_l^m(\theta, \varphi)\text{,}
\end{equation}
however, they are perhaps easiest to work with in Cartesian form,
\begin{equation}
 Z_{nl}^m(\vec{x}) = \sum_{\nu=0}^k q_{kl}^\nu |\vec{x}|^{2\nu} e_l^m(\vec{x}) \text{,}
\end{equation}
where $e_l^m$ are harmonic polynomials, $2k = n - l$ and the coefficients $q_{kl}^{\nu}$ are fixed by the orthonormality relation
\begin{equation}
 \frac{3}{4\pi} \int_{|\vec{x}| \le 1} Z_{nl}^m(\vec{x}) \overline{Z_{n'l'}^{m'}(\vec{x})} r^2 \sin\theta d\vec{x} = \delta_{nn'}\delta_{ll'}\delta^{mm'}\text{,}
 \label{eq:orthonormality}
\end{equation}
the full derivation of which can be found in \cite{Canterakis1999}.
We calculate the Zernike moments, $\Omega_{nl}^m \coloneqq \frac{3}{4\pi} \braket{f | Z_{nl}^m}$, by means of a change of basis transformation from geometric moments according to the procedure outlined by \citet{Novotni2004}.
Briefly, the Zernike moments are expressed as a linear combination of geometric moments, $m_{stu}$ (\cref{eq:geom_moms})
\begin{equation}
 \Omega_{nl}^m = \frac{3}{4\pi} \sum_{s + t + u \le n} \overline{\chi_{nlm}^{stu}} m_{stu}\text{,}
 \label{eq:zernike_from_geometric}
\end{equation}
where $\chi_{nlm}^{stu}$  are given by
\begin{equation}
\begin{split}
 \chi_{nlm}^{stu} &= c_l^m 2^{-m}%
 \sum_{\nu = 0}^l q_{kl}^\nu%
 \sum_{\alpha=0}^\nu %
 \binom{\nu}{\alpha} %
 \sum_{\beta=0}^{\nu - \alpha} %
 \binom{\nu - \alpha}{\beta} %
 \sum_{u = 0}^m (-1)^{m - u}\\
 &\quad \cdot \binom{m}{u} i^u%
 \sum_{\mu=0}^{\floor{\frac{l - m}{2} }}%
 (-1)^\mu 2^{-2\mu} %
 \binom{l}{\mu} \binom{l - \mu}{m + \mu} %
 \sum_{\nu = 0}^\mu \binom{\mu}{\nu}\text{.}
\end{split}
\end{equation}
A similar change of basis transformations for going from \gls{mtp} geometric tensors to \gls{ace} spherical tensors can be found in Appendix A. of \cite{Drautz2020}.
The main advantage of starting with geometric moments is computational speed, particularly as $\chi_{nlm}^{stu}$ need only be calculated once and can then be cached or even stored offline for reuse.

From the Zernike moments we use an independent set of 117 rotation invariants up $n = 7$ generated by the procedure in \cref{sec:poly_deps}.
These invariants are brought together in a single fingerprint vector, $\vec{\Phi}$.
Derivatives are available for both positions and weights as these are important for reconstruction.

\subsection{On completeness}

The topic of completeness of atomic descriptors has received a lot of attention (see e.g. \cite{Bartok2013a,VonLilienfeld2015,Pozdnyakov2020a,Kocer2020,Musil2021} and it is, undoubtedly, important from the point of view that all regression schemes predicting properties of atomic systems make the assumption, explicitly or implicitly, that two atomic configurations that cannot be made to coincide by a combination of rigid body translations, rotations, and for certain properties reflections, will be mapped to different points in fingerprint space.
Naturally, when inverting rotationally invariant fingerprints this assumption is equally important as, typically, we rely on the fact that up to global rotation there is a single atom density that explains a given fingerprint.

For the types of descriptors that are the focus of this work (those based on expanding an atom density in a chosen basis followed by computing rotation invariants from the resulting tensors) we can break our analysis of completeness down by considering each of the steps individually.
If the output of one step is not unique with respect to the inputs then, necessarily, the descriptor is not complete.

Provided no two atoms have identical coordinates, then the mapping of atoms onto localised feature functions is unique.
This is true even for coincident atoms so long as fixed weights are used, however, with variable weights degeneracies can occur which would require additional fingerprints to resolve.

In \cref{sec:intro} we deliberately restricted our focused to orthonormal basis functions as these are both complete and convenient for performing synthesis.
However, a set of functions need not be orthogonal to be complete in the sense of being able to approximate any function in $L^2$ arbitrary closely.
More formally, a basis can be said to be complete \citep{Courant1989} over the interval $(a, b)$ if there exists a series
\begin{equation}
 \tilde{f}(\vec{x}) = \alpha_1 \Psi_1(\vec{x}) + \alpha_2 \Psi_2(\vec{x}) + \cdots + \alpha_{n_\text{max}} \Psi_{n_\text{max}}(\vec{x})
\end{equation}
such that for every $\epsilon > 0$
\begin{equation}
 \int_a^b |f(\vec{x}) - \tilde{f}(\vec{x})|^2 d \vec{x} < \epsilon.
\end{equation}
Where the basis consists of polynomials, completeness is given by the Weierstrass approximation theorem which applies to both the geometric and Zernike moments used in this work.
Furthermore, any linear transformation of a complete basis is, itself, complete whether or not the basis is orthogonal \citep{Clement1963}.

Lastly, let us consider the completeness of the invariants themselves.
A necessary condition for a set of invariants to uniquely map moments onto a point in fingerprint space is that there be at least as many algebraically independent invariants as the number of moments minus the number of symmetrised degrees of freedom.
Working with a spherical harmonic expansion where $0 \le n   \le n_\text{max}$, $0 \le l \le l_\text{max}$ and $-l \le m \le l$ there are $n_\text{max} l_\text{max} (l_\text{max} + 2)$  moments.
\citet{Bandeira2017} have shown that with three or more radial function (i.e. $n_\text{max} \ge 3$) it is sufficient to go up to $\nu = 3$ invariants to reach independence.
While in this work, we use the reduced basis
\begin{equation}
 \sum_{n = 0}^{n_\text{max}} \sum_{l = 0}^n \sum_{m = -l}^{l} c_{nl}^m R_{nl}(r) Y_l^m(\theta, \psi)
\end{equation}
where $(n - l)$ must be even.
This requires going up to $\nu = 4$ to arrive at enough invariants.

\begin{figure}
 \includegraphics[width=0.6\columnwidth]{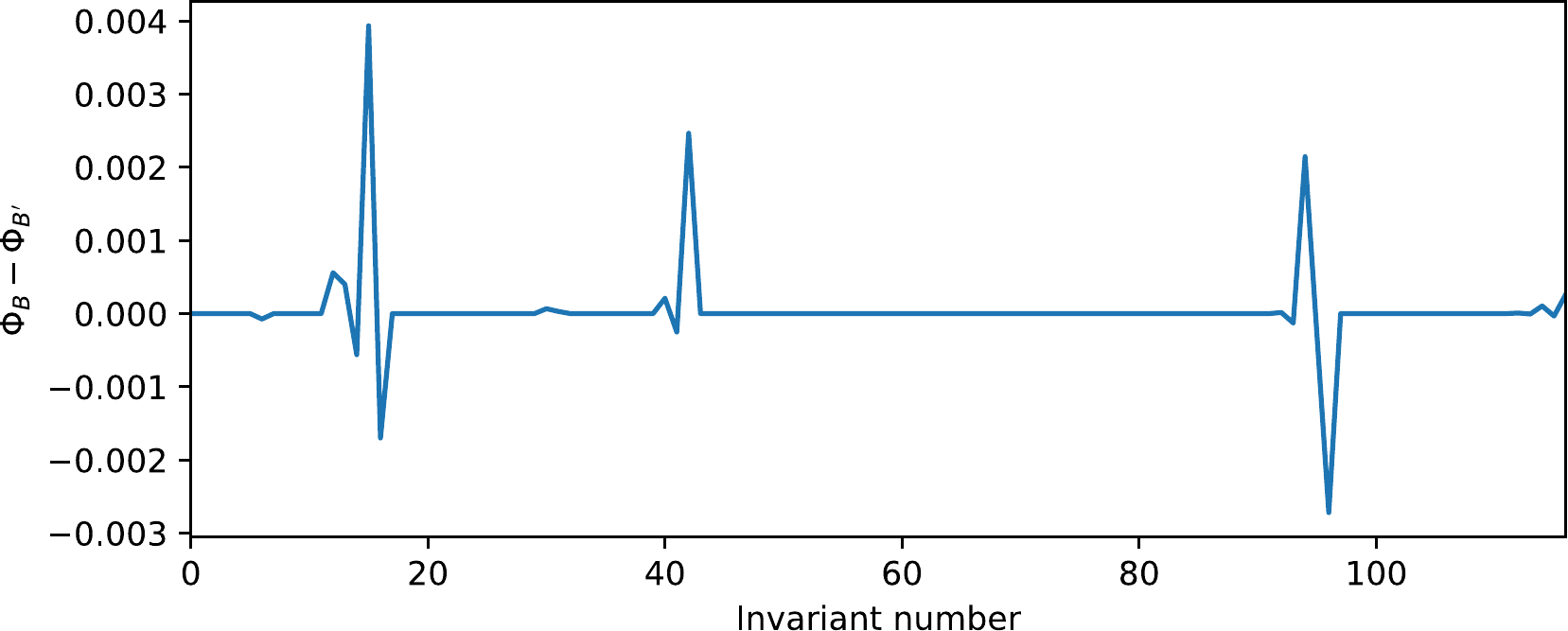}
 \caption{Difference in fingerprint vector components between atomic arrangements (labelled $B$ and $B'$) that cannot be distinguished when using $\nu = 2$ and $\nu = 3$ invariants.  All of the non-zero values correspond to $\nu = 4$ invariants.}
 \label{fig:completeness_diff}
\end{figure}

While having an algebraically complete set of invariants (or superset thereof) is necessary to distinguish all image functions, it is not sufficient as shown recently by \citet{Pozdnyakov2020a}.
In this work they provide a pair of environments that cannot be distinguished by the bispectrum (i.e. $\nu = 3$) descriptor.
In figure \cref{fig:completeness_diff} we plot the difference in fingerprint vectors generated by the MILAD descriptor for these environments.
The non-zero contributions come exclusively from 21 out of the 29 $\nu = 4$ invariants.
The fact that MILAD can distinguish these environments is, however, no guarantee that two environments could not be found that require a higher correlation order to be distinguished.

It remains an open question as to whether there is some maximum bound on $\nu$ such that all atomic configuration could be distinguished or conversely if, in the general case, it is unbounded.
As such, currently no formal statement about the injectivity of the map of moments to fingerprint used in this work can be made, however, results from \citeauthor{Pozdnyakov2020a} suggest that such examples are rare in practice.
The recent work of \citet{Pozdnyakov2021} gives a further in-depth analysis of some of the causes, and consequences, of incompleteness in rotationally-invariant atomic descriptors.

Finally, assuming a certain minimum interatomic distance, the number of invariants needed to uniquely describe an atomic environment will naturally increase with the number of atoms, $N$.
By considering that there are $3N - 3$ degrees of freedom in a rotation symmetrised environment we would expect to need at least this many invariants.
In reality, the number needed is likely to be greater owing to the fact the the types of invariants discussed in this work are capable of describing functions that fall outside of the limited class of atom densities considered.
One numerical way to probe this question more generally would be with the use of an autoencoder.
By successively decreasing the number of artificial neurons in the bottleneck, one could look for a signature, discontinuous, increase in the decoding RMSD over a large set of examples environments.
This will be explored as part of future work, however we briefly revisit this issue when discussing results from reconstruction experiments further below.

\section{Reconstruction}
\label{sec:reconstruction}

The goal of reconstruction is to start with a set of moment invariants and general information about the set of feature functions used (function type, range of weights, etc.) and reproduce the original set of atomic positions, and optionally, their atomic species.
It is possible to start with semi-random configurations of atoms and perform a global optimisation of the residuals with respect to the known invariants (similar to random structure searching \citep{Pickard2011}), however this scales poorly as the number of atoms grows due to the large number of local minima encountered.
On the other hand, performing a local minimisation of the atomic coordinates with respect to known moments is significantly more reliable.

The problem of retrieving a set of moments starting from rotation invariants has commonalities with the more well known phase retrieval problem that lies at the heart of solving for crystal structures from their x-ray diffraction pattern.
In this case, the symmetrisation is over the translation group which manifests in only the intensities of the diffraction peaks being observed, preventing the structure from being solved directly by simply performing the inverse Fourier transform due to the missing phases.
One of the most well known methods for phase retrieval is the Gerchberg-Saxton algorithm \citep{Gerchberg1972} which also has a similar, iterative optimisation, structure to that used here.
The algorithm starts with random initial phases and proceeds by iteratively performing forward and backwards Fourier transforms, modifying the phases in-between to better match the observed diffraction pattern until the difference falls below a given threshold.
The structure of the orientation retrieval problem being addressed here differs primarily by having to simultaneously solve for a hierarchy of many-point correlations (up to the maximum correlation order used), whereas phase retrieval typically deals with two point correlations (from the square of the Fourier transform).

Another closely related problem is that of multi-reference alignment, commonly used to solve the structure of single molecules in cryogenic electron microscopy (cryo-EM).
Here an image is taken of many copies of the same molecule with different orientations from which the three-dimensional structure is determined.
One of the more relevant algorithms from this community can be found in the work of \citet{Bandeira2017}.
The authors give a closed-form procedure to recover the spherical harmonic expansions coefficient from a set of $\nu = 2$ and $\nu = 3$ invariants.
The algorithm is included in our codebase and has been tested on a number of atomic systems.
The major shortcoming is that the procedure only works for, so called, generic signals and fails in the presence of symmetry.
This is because symmetry causes some of the invariants to be trivial leading to a situation where a linear system that must be solved as part of the algorithm becomes underdetermined.
By contrast, the numerical algorithm presented here relies on finding a least-squares solution which is biased by the prior-knowledge we have about the class of signals making it more robust to such situations.

\subsection{Moments from invariants}
\label{sec:moms_from_invs}

\begin{tikzpicture}[
  scale=0.8,
  auto, 
  transform shape,
  block/.style={draw, align=center, minimum width=5em, minimum height=3.1em},
  annotation/.style={align=center, font=\tiny},
  line join=round,font=\sffamily,3d cube/.cd, num cubes x=1,num cubes y=1,num cubes z=1
  ]

\tdplotsetmaincoords{70}{45}
\begin{scope}[xshift=0cm, scale=0.42, tdplot_main_coords,local bounding box=moments]
  \path pic{cube array={num cubes x=3,num cubes y=3,num cubes z=3}};
\end{scope}
\tdplotsetmaincoords{85}{190} 
\begin{scope}[xshift=4cm, scale=0.42, tdplot_main_coords,local bounding box=invariants]
  \path pic{cube array={num cubes x=5}};
\end{scope}

Inverse links
\draw[->,ultra thick,draw=gray] ($(invariants.west) + (-0.4em, 0)$) -- ++(-1.5cm, 0);

\begin{scope}[yshift=-1.4cm]
 \node (moments_label) at (0, 0) {Moments};
 \node[anchor=north] at (moments_label.south) {$c_{nl}^m, m_{stu}, \Omega_{nl}^m$};
 \node (invariants_label) at (4, 0) {Fingerprints};
 \node[anchor=north] at (invariants_label.south) {$\vec{\Phi}$};
\end{scope}
\end{tikzpicture}

The invariants, $\Phi_i$, form a latent space from which the corresponding moments can be recovered by solving the system of polynomials which couples the two.
The system of equations is underdetermined by three equations corresponding to the missing orientation information; a degeneracy that will be resolved during reconstruction, however the new orientation will be uncorrelated with the original.
The zeroth-order moment is itself a rotation invariant and encodes the total mass of the environment, in the case of geometric moments $m_{000} = \sum_i^{N_A} w_i$ while for Zernike moments $\Omega_{00}^0 = \frac{3}{4 \pi} m_{000}$.
The invariant of first order moments
\begin{equation*}
 \Phi_1 = \frac{1}{\sqrt{3}} \left( 2 {c}_{11}^{-1} {c}_{11}^1 - {{c}_{11}^0}^{2} \right)\text{,}
\end{equation*}
encodes the distance of the centre of mass from the origin, valid solutions for which can be found trivially.
From here it is possible to solve the system of equations 
iteratively, where at each iteration we include all invariants containing moments up to a maximum angular frequency, $l^\prime$, effectively building up a more and more detailed reconstruction of the original environment.
At the end of each iteration, $l^\prime$ is incremented by one until we reach $l_\text{max}$.
This procedure can encounter local minima and we therefore make several attempts (typically up to two at each $l^\prime$) to find a solution until the root-mean-square deviation from the known invariants drops to below a given threshold, typically taken to be $10^{-5}$.
The orientation recovered is random, being determined by arbitrary choices made in picking a particular solution for the first and second order moments (or when solving for higher order moments if the original environment is highly symmetric).
We use the Levenberg-Marquardt \citep{More1978} least-squares solver implemented in SciPy \cite{Virtanen2020} to perform this procedure, starting with an initial set of moments created from an environment of atoms placed randomly within the cutoff sphere and with species chosen randomly from the full set supported by the descriptor.
However, we find that the solution is not particularly sensitive to the starting point and even random moments will often converge successfully.
\Cref{algo:moments_finding} gives the pseudocode for the entire procedure and the python code is also available \footnote{\url{https://github.com/muhrin/milad/blob/v0.3.0/milad/optimisers/moments_optimiser.py}}.

Once all of the moments are found, an approximation of the original function, $f(\vec{r})$, can be reconstructed (\cref{eq:reconstruction}).
\begin{figure*}[ht]
 \includegraphics[width=0.98\textwidth]{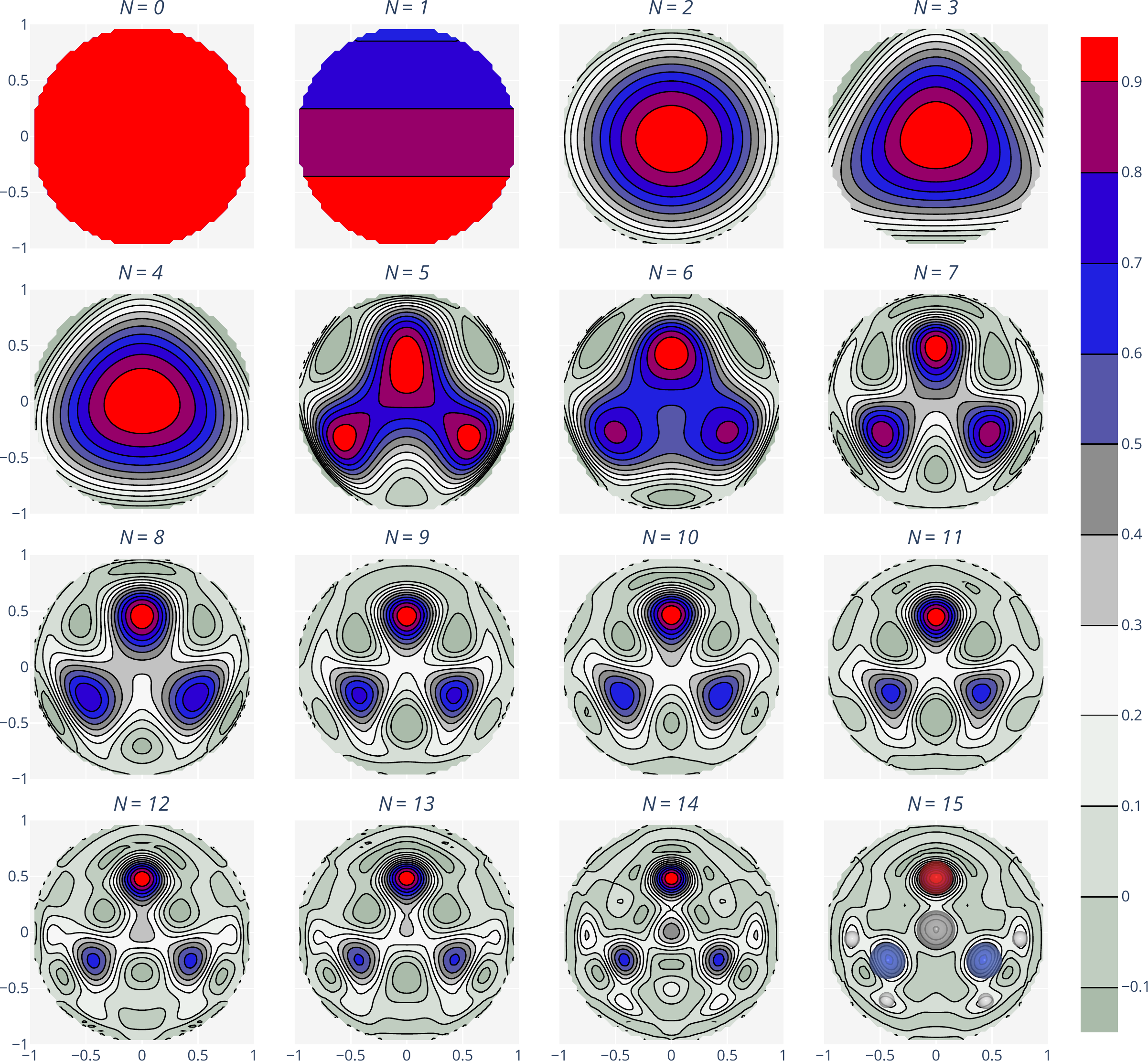}
 \caption{Reconstructions from Zernike moments with increasing maximum expansion order $N$.  The contour data is taken as a slice through a 3D grid corresponding to the plane of the molecule, a reduced opacity version of which is shown over $N = 15$.  Atomic species have been mapped to the following weights of delta functions: $w_\text{H} = 1.125, w_\text{C} = 2.375, w_\text{N} = 3.625, w_\text{O} =  4.875$.
 The fractional colour scale has been chosen to correspond to the standard CPK colours for each element.}
 \label{fig:urea_reconstructions}
\end{figure*}
\Cref{fig:urea_reconstructions} shows a urea molecule reconstructed using various values of maximum expansion order.
\begin{algorithm}[H]
 \SetKwComment{tcp}{\# }{}

 \SetKwInOut{Inputs}{Inputs}
 \SetKwInOut{Outputs}{Outputs}

 \SetKwData{Moments}{$c_{nl}^m$}
 \SetKwData{PartialMoments}{$c_{nl}^m[l \le l']$}
 \SetKwData{Fingerprint}{$\Phi_i$}
 \SetKwData{PartialFingerprint}{$\Phi_i{[l_\text{max} \le l']}$}
 \SetKwData{Point}{point}
 \SetKwData{Positions}{positions}
 \SetKwData{Retries}{max\_retries}
 \SetKwData{rmsd}{rmsd}
 \SetKwData{BestRmsd}{best\_rmsd}
 \SetKwData{BestMoments}{best\_moments}
 \SetKwData{Tol}{tol}
 \SetKwData{Moms}{moments}
 
 \SetKwFunction{CalculateGrid}{CalculateGrid}
 \SetKwFunction{CalculateMoments}{CalculateMoments}
 \SetKwFunction{Append}{append}
 \SetKwFunction{InitialConfiguration}{RandomInitialConfiguration}
 \SetKwFunction{LstSq}{LstSq}
 \SetKwFunction{Randomise}{Randomise}

 \Inputs{%
 \Fingerprint: fingerprint vector,
 $N_A$: num. atoms,
 $r_{cut}$: the descriptor cutoff radius,
 $r_\text{min}$: minimum atoms separation,
 \Tol: the least-squares tolerance on the RMSD,
 \Retries: max retries for each $l$%
 }
 \Outputs{$c_{nl}^m$ the moments}
 
 \tcp{Make an initial guess and calculate the corresponding moments}
 \Positions $\gets$ \InitialConfiguration($N_A$, $r_\text{cut}$, $r_\text{min}$)\;
  \Moments $\gets$ \CalculateMoments(\Positions)\; 

\For{$l' \gets 1$ \KwTo $l_\text{max}$}{
  \BestRmsd $\gets \infty$\;
  \For{$1$ \KwTo $\Retries$}{
    \tcp{Gradient descent w.r.t. moments}
    \Moms, \rmsd $\gets$ \LstSq(\PartialMoments, \PartialFingerprint)\;
    \If{\rmsd $<$ \BestRmsd}{
        \tcp{Retain this as the best minimum found so far}
        \BestRmsd $\gets$ \rmsd\;
        \BestMoments $\gets$ \Moms\;
    }
    \eIf{\BestRmsd $<$ \Tol}{
        break\;
    }{
        $c_{nl}^m [l = l'] \gets$ \Randomise($c_{nl}^m [l = l']$)\;
    }
  }
  \PartialMoments $\gets$ \BestMoments\;
}
\Return \Moments\;
 \caption{Numerical algorithm to solve for a set of moments from a given fingerprint.  We use square brackets to indicate that only a subset of the array matching the condition is being used.}
 \label{algo:moments_finding}
\end{algorithm}

\subsection{Atoms from moments}
\label{sec:atoms_from_moments}

\begin{tikzpicture}[
  scale=0.8,
  auto, 
  transform shape,
  block/.style={draw, align=center, minimum width=5em, minimum height=3.1em},
  annotation/.style={align=center, font=\tiny},
  line join=round,font=\sffamily,3d cube/.cd, num cubes x=1,num cubes y=1,num cubes z=1
  ]
  
\tdplotsetmaincoords{85}{190} 

\node (env) at (0, 0) {\includegraphics[width=2cm]{urea_3d.png}};
\begin{scope}[xshift=4cm,scale=1]
 \node[minimum width=2.2cm, minimum height=1.36cm] (features) {};
 \fill[draw=white,outer color=white,inner color=black!60!black,minimum width=3cm] (0.05,0.04) circle (0.15cm);
 \fill[draw=white,outer color=white,inner color=black!60!black,minimum width=3cm] (0.46,-0.24) circle (0.15cm);
 \fill[draw=white,outer color=white,inner color=black!60!black,minimum width=3cm] (0.32,-0.38) circle (0.15cm);
 \fill[draw=white,outer color=white,inner color=black!60!black,minimum width=3cm] (0.88,-0.04) circle (0.15cm);
 \fill[draw=white,outer color=white,inner color=black!60!black,minimum width=3cm] (-0.86,-0.04) circle (0.15cm);
 \fill[draw=white,outer color=white,inner color=black!60!black,minimum width=3cm] (-0.65,-0.51) circle (0.15cm);
 \fill[draw=white,outer color=white,inner color=black!60!black,minimum width=3cm] (-0.56,-0.12) circle (0.15cm);
 \fill[draw=white,outer color=white,inner color=black!60!black,minimum width=3cm] (0.18,0.4) circle (0.15cm);
\end{scope}

\tdplotsetmaincoords{70}{45}
\begin{scope}[xshift=8cm, scale=0.42, tdplot_main_coords,local bounding box=moments]
  \path pic{cube array={num cubes x=3,num cubes y=3,num cubes z=3}};
\end{scope}

\draw[<-,ultra thick] ($(env.east) + (0.2, 0)$) --  ($(features.west) + (-0.2, 0)$);
\draw[<-,ultra thick,draw=gray] ($(features.east) + (0.2, 0)$) -- ++(1.5cm, 0);

\begin{scope}[yshift=-1.4cm]
 \node (envs_label) {Environments};
 \node[anchor=north] at (envs_label.south) {$\{\vec{r}_i, Z_i, \ldots \}$};
 \node (features_label) at (4, 0) {Features};
 \node[anchor=north] at (features_label.south) {$f(\vec{r}_i, Z_i, \ldots)$};
 \node (moments_label) at (8, 0) {Moments};
 \node[anchor=north] at (moments_label.south) {$c_{nl}^m, m_{stu}, \Omega_{nl}^m$};
\end{scope}

\end{tikzpicture}

Next, we are tasked with recovering the atomic positions and species from the moments.
Once again this can be achieved by means of local minimisation, this time between the atomic degrees of freedom and the moments themselves (for the sake of this procedure the features can be largely ignored as there is an analytic expression between them and the atomic coordinates).

If all of the weights are the same (i.e. all atoms are of the same specie) then the number of atoms can be determined directly from $c_{00}^0$, otherwise this moment simply acts as a constraint on the total weight of the delta (or Gaussian) functions in the environment, an alternative is to use two vectors of invariants, one encoding only positions and the other species, similar to the scheme proposed by \citet{Artrith2017}.
If no detailed information about the number of atoms of each specie is known there are several ways to proceed,
\begin{enumerate*}[label=\arabic*)]
 \item attempt multiple optimisations with different numbers of atoms (that are consistent with $c_{00}^0$ and the allowed range of feature function weights) and keep the best fit, or,
 \item use more atoms than necessary and merge overlapping atoms by summing their weights as part of post-processing.
\end{enumerate*}

\begin{figure*}[ht]
 \includegraphics[width=0.85\textwidth]{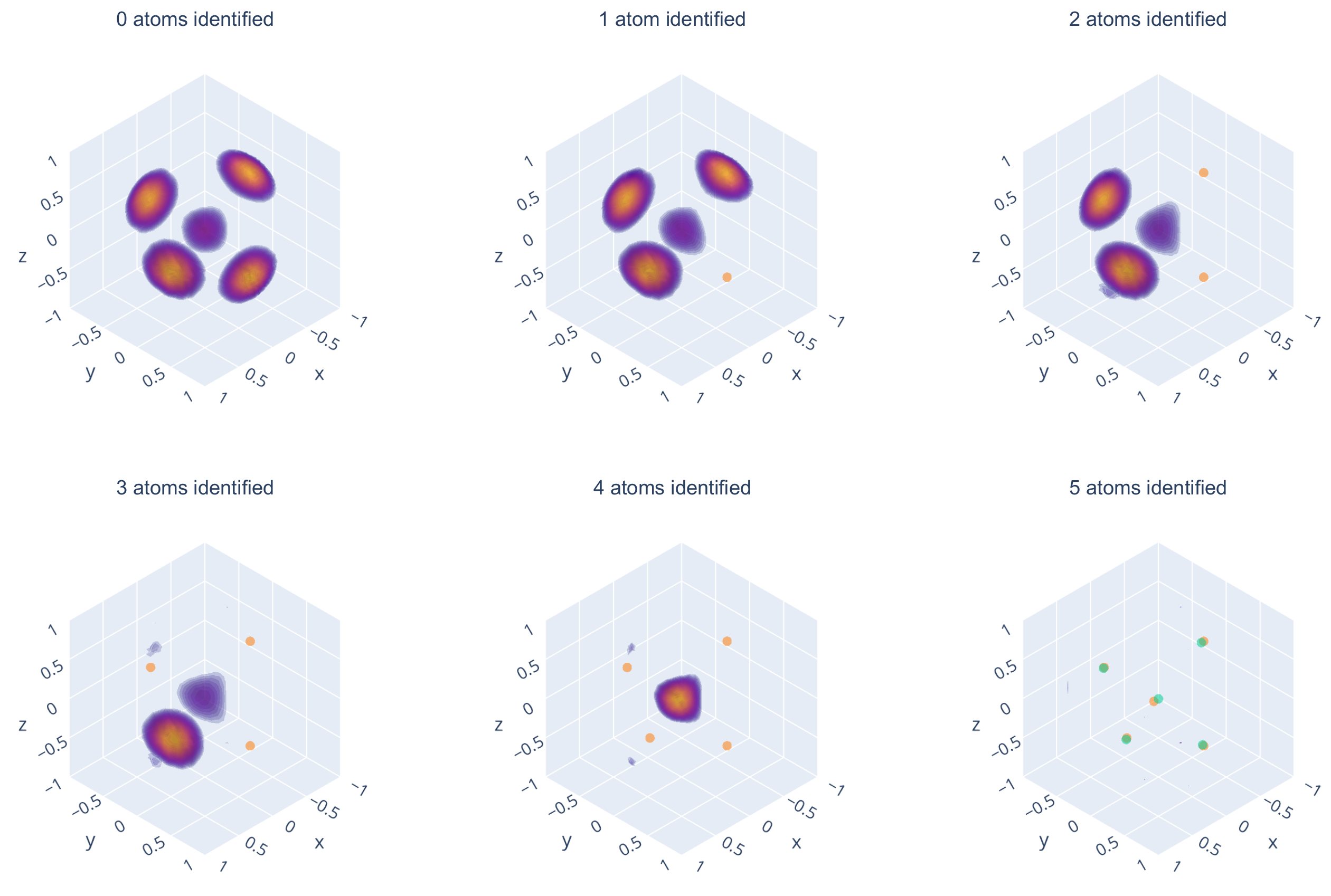}
 \caption{Peak finding from an approximation of the original environment, $\tilde{f}$, based on Zernike moments.  Atoms are located one by one whereupon they are replaced by an orange marker and the signal is subtracted for the next iteration.  The final panel shows the original locations of the atoms as green markers.}
 \label{fig:atoms_reconstructions}
\end{figure*}

The initial atomic positions for the optimisation are chosen using peak finding where we iteratively locate the highest density point in a 3D grid whose values are calculated according to \cref{eq:reconstruction}.
A cubic grid is used that contains the cutoff sphere, typically sampled using $31^3$ points.
Once an atom is found we
\begin{enumerate}
  \item subtract a single-atom signal from the density grid, and,
  \item zero out all grid values in the vicinity of the located atom to ensure the algorithm does not place any two atoms too close.
\end{enumerate}
The procedure is summarised in \cref{algo:peak_finding} and an example is shown in \cref{fig:atoms_reconstructions}.

\begin{algorithm}[H]
 \SetKwComment{tcp}{\# }{}

 \SetKwInOut{Inputs}{Inputs}
 \SetKwInOut{Outputs}{Outputs}

 \SetKwData{Grid}{grid}
 \SetKwData{Values}{values}
 \SetKwData{Point}{point}
 \SetKwData{Positions}{positions}
 \SetKwData{Minsep}{$r_\text{min}$}
 \SetKwData{V}{v}
 \SetKwData{P}{p}
 
 \SetKwFunction{CalculateGrid}{CalculateGrid}
 \SetKwFunction{CalculateMoments}{CalculateMoments}
 \SetKwFunction{argmax}{argmax}
 \SetKwFunction{argwhere}{argwhere}
 \SetKwFunction{Max}{max}
 \SetKwFunction{Append}{append}

 \Inputs{%
 $c_{nl}^m$: moments,
 $N_A$: num. atoms,
 \Minsep: minimum atoms separation,
 \Grid: the array of Cartesian coordinates to perform reconstruction over
 }
 \Outputs{The atomic positions}
 
 \Positions $\gets$ list()\;
 \tcp{Get values of function reconstruction on grid}
 \Values $\gets$ \CalculateGrid(\Grid, $c_{nl}^m$)\; 
\For{$i \gets 1$ \KwTo $N_A$}{
  \tcp{Get grid values for isolated atom at\\ current maximum value}
  \Point $\gets$ \Grid[\argmax(\Values)]\;
  \V $\gets$ \CalculateGrid(\CalculateMoments(\Point))\;
  \tcp{Remove contribution of this atom}
  \Values $\gets$ \Values $-$ \V\;
  \tcp{Zero grid values near this point\\so no further atoms are placed near}
  \Values[\argwhere(grid - \Point $ < $ \Minsep)] = 0\;
  \Positions.append(\Point)\;
}
\Return \Positions\;
 \caption{Peak finding algorithm for determining discreet atom positions from the moments.}
 \label{algo:peak_finding}
\end{algorithm}

Finally, to ensure that the optimisation avoids solutions that place atoms closer than is physically reasonable we add the following energy term to the cost function that biases it away from these configurations.
\begin{equation}
 E(r_{ij}) = \begin{cases}
         4 \epsilon \left[ \left( \frac{\sigma}{r_{ij}} \right)^n + k_1 r_{ij} + k_2 \right] & \text{if $r_{ij} < r_\text{min}$}\\
         0 & \text{otherwise}
        \end{cases}
\end{equation}
where $r_{ij}$ is the interatomic separation, $k_1$ and $k_2$ are constants set such that $E(r_{ij})$, and $\frac{dE}{dr_{ij}}$ both approach zero smoothly at the cutoff.  
We find that $\epsilon = 0.1, \sigma = 1, n = 2$ and $r_\text{min} = 0.55\text{\AA}$ work well and use these values throughout.
The total loss function is thus
\begin{equation}
 L \left(c_{nl}^m, c_{nl}^{\prime m}, \{\vec{r}_1, \ldots, \vec{r}_{N_A}\} \right) = \text{RMSD}(c_{nl}^{\prime m}, c_{nl}^m) + \sum_{ij, i \ne j}^{N_A} E(\lVert \vec{r_j} - {\vec{r}_i} \rVert),
\end{equation}
where $c_{nl}^{\prime m}$ are the moments calculated from the atomic configuration at each step during the optimisation.

\subsection{Putting it all together}
\label{sec:reconstruction_all}

Even for the case of noise-free fingerprints the above procedures are not guaranteed to find the original atomic environment.
The primary reason for this is that both optimisations (moments from invariants and atoms from moments) are performed over non-convex functions and, as such, they can end up in the wrong minimum.
This situation is easily detected as the gradients vanish while the RMSD remains high.

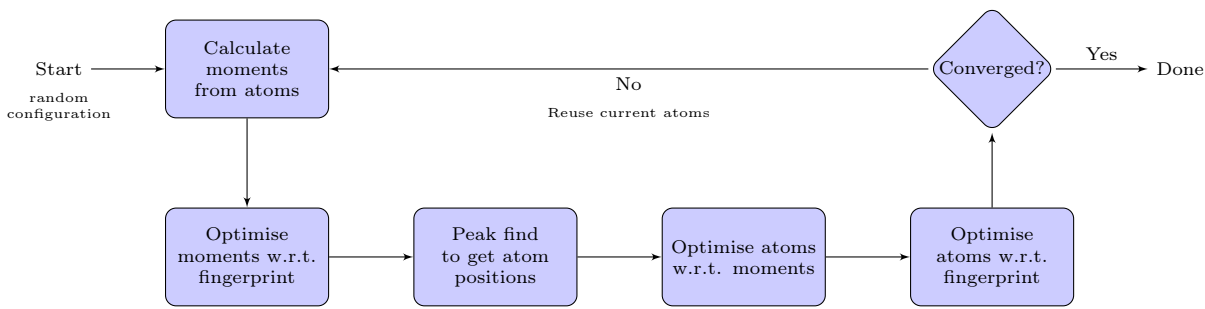
\begin{figure*}[ht]
\begin{center}
\tikzstyle{decision} = [diamond, draw, fill=blue!20, 
    text width=4.5em, text badly centered, node distance=3cm, inner sep=0pt, rounded corners]
\tikzstyle{block} = [rectangle, draw, fill=blue!20, 
    text width=6em, text centered, rounded corners, minimum height=4em]
\tikzstyle{line} = [draw, -latex']
\begin{tikzpicture}[node distance = 3.3cm, auto, font=\scriptsize]
    \node (start) {Start};
    \node[anchor=north, font=\tiny, align=center] at (start.south) {random\\configuration};
    \node[block, right of=start, node distance=2.5cm] (calc-moms) {Calculate moments from atoms};
    \node[block, below of=calc-moms, node distance=2.5cm] (opt-moms) {Optimise moments w.r.t. fingerprint};
    \node[block, right of=opt-moms] (find-atoms) {Peak find to get atom positions};
    \node[block, right of=find-atoms] (opt-atoms-moms) {Optimise atoms w.r.t. moments};
    \node[block, right of=opt-atoms-moms] (opt-atoms) {Optimise atoms w.r.t. fingerprint};
    \node[decision, above of=opt-atoms, node distance=2.5cm] (converged) {Converged?};
    \node[right of=converged, node distance=2.5cm] (done) {Done};

    \path [line] (start) -- (calc-moms);
    \path [line] (calc-moms) -- (opt-moms);
    \path [line] (opt-moms) -- (find-atoms);
    \path [line] (find-atoms) -- (opt-atoms-moms);
    \path [line] (opt-atoms-moms) -- (opt-atoms);
    \path [line] (opt-atoms) -- (converged);
    \draw [line] (converged) -- (calc-moms) node[midway] (no) {No};
    \path [line] (converged) -- node {Yes} (done);
    
    \node [anchor=north, font=\tiny, align=center] at (no.south) {Reuse current atoms};
\end{tikzpicture}

\end{center}
\caption{Iterative scheme that finds atomic positions, and optionally species, by alternating between an optimisation of the moments and an optimisation of the atomic positions, both with respect to the known fingerprints.}
\label{fig:iterative_inversion}
\end{figure*}

To overcome this we implement an algorithm according to the scheme shown in \cref{fig:iterative_inversion} which alternates between an optimisation of the moments followed by a optimisation of the corresponding atomic degrees of freedom.
This is repeated a number of times until the RMSD drops below a preset threshold.

The algorithm presented here is designed to reconstruct a single atomic environment and makes the assumption that all atoms lie within the corresponding cutoff sphere.
Outside this domain, the basis is no longer orthogonal, and therefore it is not strictly possible to reconstruct the atom density according to \cref{eq:reconstruction}.
Instead, the reconstruction of an atomic structure consisting of multiple overlapping cutoff spheres could be performed by simultaneously minimising the loss over all environments in each step.
However, this greatly complicates the problem as now the orientations and positions (if not fixed) of overlapping environments are coupled, potentially introducing additional minima to the optimisation landscape.
Furthermore, if the environments are not all identical, the problem is no longer permutationally invariant with respect to the assignment of fingerprints to each environment.
For these reasons, we leave the reconstruction of multi-centre configurations for future work.

\section{Experiments}
\label{sec:experiments}

\subsection{Reconstruction}

To assess the quality of various reconstructions from moment invariants we use a subset of the QM9 database \citep{Ruddigkeit2012,Ramakrishnan2014}, a database of small organic molecules.
Three molecules of each size (in number of atoms) were chosen randomly, ranging from $N_A = 3 \text{ to } 29$, with the exception of 3 atom molecules of which there are only two and 28 atom molecules of which there are none.
This corresponds to a total test set of 77 molecules.
The full set of QM9 IDs used can be found in the Supplemental Material \cite{SI}.
We use weighted delta functions as features.
If only position information is being used, then the weights are all fixed to 1, otherwise the atomic numbers of HCONF are mapped onto the continuous range $1 \to 2$ (i.e. $w_\text{H} = 1.1, w_\text{C} = 1.3, w_\text{O} = 1.5, w_\text{N} = 1.7, w_\text{F} = 1.9$).
In the latter case the weights are free to vary during optimisation, and are mapped back onto the correct integers as part of post-processing. 
For example, a value falling in the interval $[1, 1.2)$ would be mapped to hydrogen.
The cutoff sphere is set at 5\AA{} such as to accommodate the largest molecule in the data set.
Each molecule is positioned by finding the smallest bounding sphere containing all atoms using the miniball library \footnote{\url{https://pypi.org/project/miniball/}} and translating the centre of the sphere to the origin.

In the first instance, the quality of the reconstruction is judged by the root mean square of the difference between the calculated fingerprint and that of the recovered atomic environment, $\sqrt{\sum_i \left( \Phi_i - \Phi'_i \right)^2 / N_\Phi}$.
The validity of this as a measure of structural similarity is demonstrated in the Supplemental Material \cite{SI}.
We note that MILAD fingerprints are unable to distinguish between chiral versions of a molecule and therefore a low RMSD can indicate that the original \text{or} a chiral counterpart have been recovered.
A summary of the parameters used is tabulated below.
\begin{center}
{\small
\setlength\extrarowheight{-10pt}
\begin{tabular}{l r}
\hline\hline
Total molecules & 77 \\
Species & HCONF \\
Feature function & delta \\
Weights & 1 $\to$ 2 \\
Cutoff & 5\AA \\
Invariants & Zernike (to $n_\text{max} = 7$), \\
 &  117 total \\
Reconstruction attempts & 3 \\
\hline\hline
\end{tabular}
}
\end{center}
In each case the fingerprint is calculated and used to perform a reconstruction according to the procedure outlined in \cref{sec:reconstruction_all}.

\begin{figure*}[!ht]
 \centering
 \includegraphics[width=\textwidth]{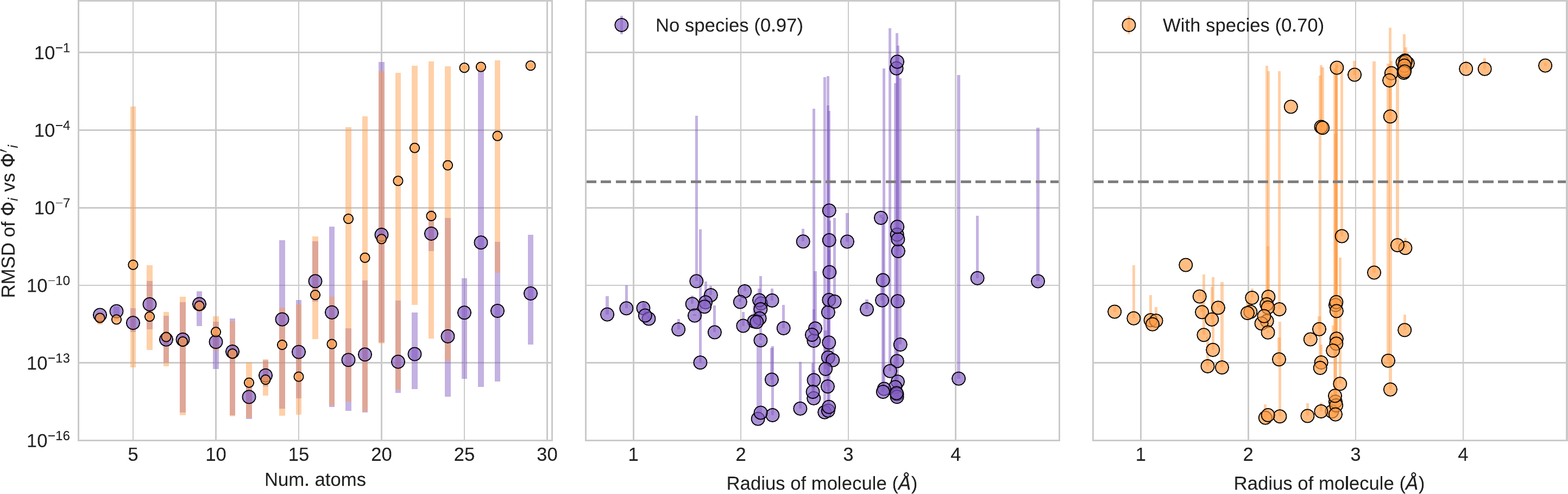}
 \caption{Results showing the RMSD of the original fingerprints versus those calculated from the reconstructed molecules.  The leftmost panel shows results aggregated by number of atoms.  The right panels shows results for each molecule individually where the marker is placed at the lowest RMSD.  Error bars indicate the min/max for that data point over several reconstruction attempts.  A value of $\approx{}10^{-6}$ (dashed line) represents a faithful reconstruction, the legend shows the proportion of markers falling below this threshold.}
 \label{fig:atoms_recovery_comparison}
\end{figure*}

\Cref{fig:atoms_recovery_comparison} shows results from the reconstruction of atomic positions (orange) and atomic positions plus species (purple).
For each molecule three reconstruction attempts were made, in each  case starting from a random configuration with the correct number of atoms.
A manual inspection of the recovered molecules indicates that an RMSD of $< 10^{-6}$ represents a visually indistinguishable reconstruction versus the original and we use this as the threshold of successful reconstruction.
The recovery of atomic positions is generally highly reliable, achieving successful reconstruction for 97\% of molecules and all molecules with a radius of less than $\approx{}3.4$\AA.
Any larger than this and the effective resolution of the basis (see \cref{fig:urea_reconstructions}) is insufficient for consistent reconstruction on every attempt.
The largest molecules that can be decoded have 29 atoms translating to 85 degrees of freedom after symmetrisation, with consistent recovery up to 19 atoms or 54 degrees of freedom.
These numbers provide some indication of the efficiency with which the fingerprint encodes the symmetrised atom density given that it consists of 117 invariants.
Reconstruction of positions and species is less reliable, achieving an overall 70\% success rate with consistent recovery up to $\approx{}2.4$\AA.
In this case there is one more degree of freedom per atom meaning that the largest molecule decoded, with 27 atoms has, 105 degrees of freedom while the largest molecules that can be consistently decoded have 17 atoms and 65 degrees of freedom.

Next, to improve upon the recovery of species information we take the reconstructed atomic positions from the last experiment and perform a reconstruction of the weights with respect to fingerprints with species information.
This represents a scenario where two fingerprint vectors are supplied, one containing position information only and the other augmented with species information, similar to using an additional colour channel in an image.

\begin{figure*}[!ht]
 \centering
 \includegraphics[width=\textwidth]{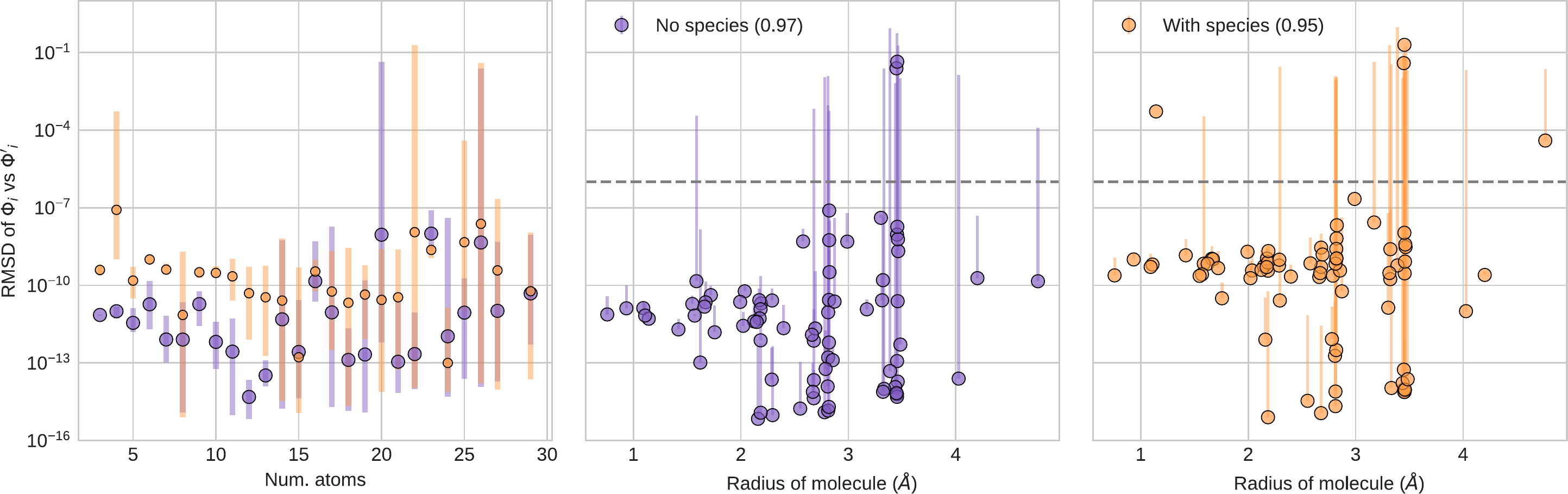}
 \caption{A plot showing recovery of atomic species using a second fingerprint vector including species information.  Results from decoding atomic positions only (`no species') are replotted from \cref{fig:atoms_recovery_comparison} for ease of comparison.}
 \label{fig:species_recovery}
\end{figure*}

\Cref{fig:species_recovery} shows the species recovery results with the original position recovery data replotted for easy comparison.
This shows a notable improvement with 95\% of structures now being recovered correctly.
With few exceptions, the structures whose positions were correctly recovered in the previous experiment now have their species correctly decoded.

\begin{figure*}[!ht]
 \centering
 \subfloat[No species]{\includegraphics[width=0.4\textwidth]{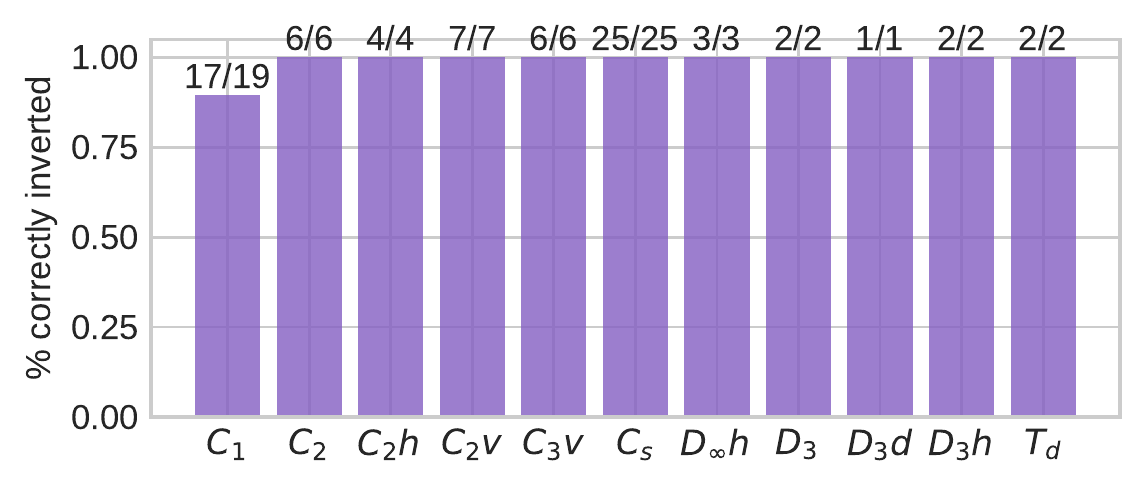}}
\subfloat[With species (single fingerprint)]{\includegraphics[width=0.4\textwidth]{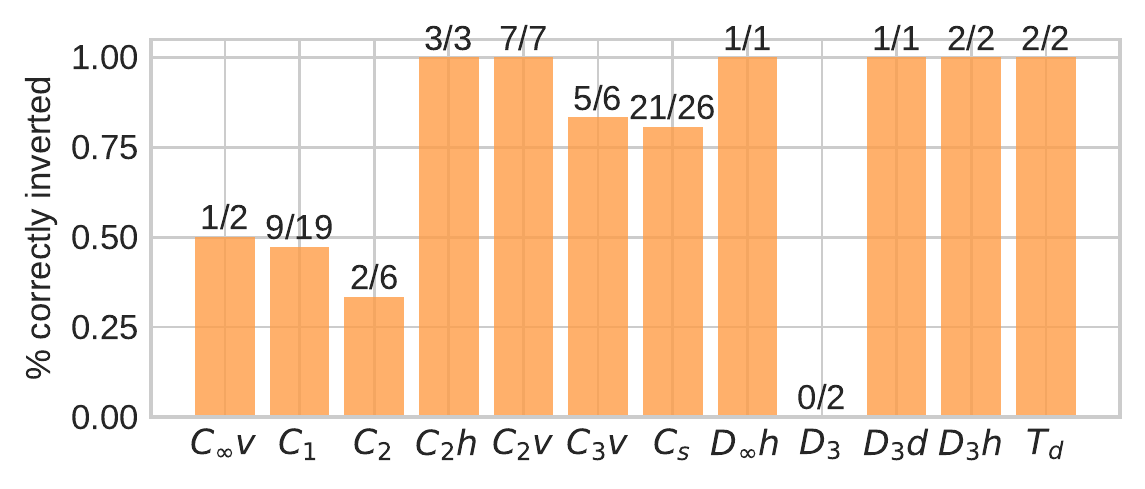}}\\
\subfloat[With species (two fingerprints)]{\includegraphics[width=0.4\textwidth]{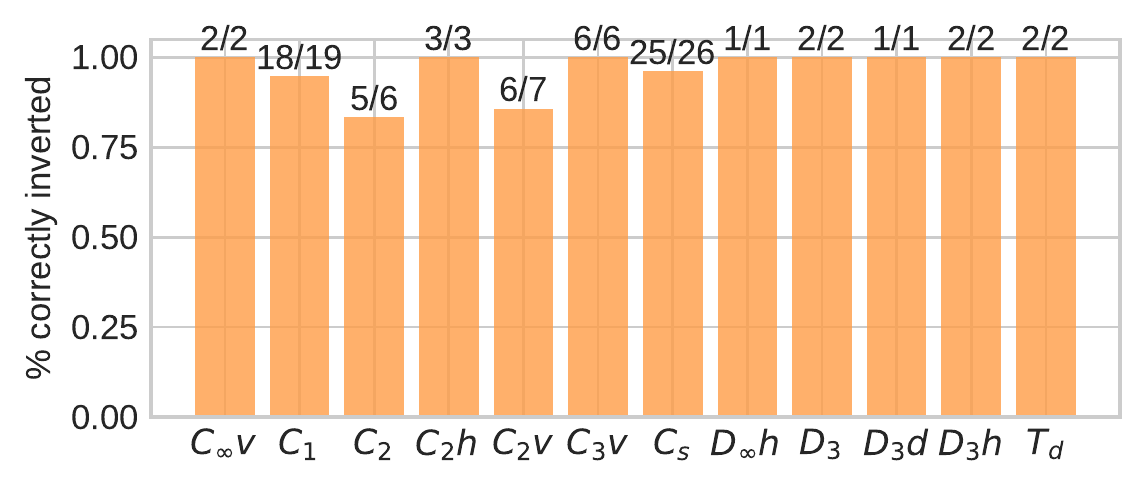}}
 \caption{Breakdown by point group showing the proportion of molecules correctly inverted.  Numbers above bars indicate the absolute number correctly inverted out of the total.}
 \label{fig:pg_analysis}
\end{figure*}

Finally, \cref{fig:pg_analysis} shows the reconstruction results broken down by the point group of the molecules.
This analysis is important to verify that the inversion algorithm works correctly in the presence of symmetry which can lead to many invariants being zero.
As can be seen, there is no discernible trend that would suggest that the algorithm cannot deal with symmetric molecules.

\subsection{Neural network potential}

\begin{figure*}[ht]
\begin{tikzpicture}
    
    \node (phi1) at (-2,-0) {$\vec{\Phi}_1$};
    \node (phi2) at (-2,-1) {$\vec{\Phi}_2$};
    \node at (-2,-1.6) {$\vdots$};
    \node (phiN) at (-2,-2.5) {$\vec{\Phi}_N$};

    \node[fill=white] (nn1) at (0, 0) {\includegraphics[width=1cm]{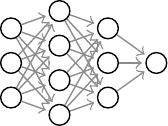}};
    \node[fill=white] (nn2) at (0, -1cm) {\includegraphics[width=1cm]{neural_network.pdf}};
    \node[fill=white] (nnN) at (0, -2.5cm) {\includegraphics[width=1cm]{neural_network.pdf}};
    
    \node (eps1) at (2,0) {$\epsilon_1$};
    \node (eps2) at (2,-1) {$\epsilon_2$};
    \node at (2,-1.6) {$\vdots$};
    \node (epsN) at (2,-2.5) {$\epsilon_N$};
    
    \node[draw] (sum) at (3.5, -1) {$\sum$};
    \node[anchor=west] (total_energy) at (4.5,-1) {$E$};
    \node[anchor=west] at (4.5, -1.6) {$\vec{F}_i = -\frac{\partial \epsilon_i}{\partial \vec{\Phi}_i} \sum\limits_{j}^{n} \frac{\partial \vec{\Phi}_i}{\partial \vec{r}_{ij}}$};

    \draw[thick] (phi1) -- (nn1.west);
    \draw[thick] (phi2) -- (nn2.west);
    \draw[thick] (phiN) -- (nnN.west);
    
    \draw[thick] (nn1.east) -- (eps1);
    \draw[thick] (nn2.east) -- (eps2);
    \draw[thick] (nnN.east) -- (epsN);
    
    \draw[thick] (eps1) -| (sum);
    \draw[thick] (eps2) -- (sum);
    \draw[thick] (epsN) -| (sum);
    
    \draw[thick] (sum) -- (total_energy);
    
    \draw[thick] (-4.7cm, -0.5cm) |- (phi1);
    \draw[thick] (-3.3, -1.45) |- (phi2);
    \draw[thick] (-5, -1.3) |- (phiN);

    \node[inner sep=0pt,anchor=east] (urea) at (-3,-1.4)
    {\includegraphics[width=4.1cm]{urea_3d.png}};

    \draw[dashed] (-5, -1.32) circle (1.3cm);
    \draw[->, thick] (-5, -1.32) -- node[pos=0.7,sloped,anchor=north] {\scriptsize $r_\text{cut}$} +(30:1.3cm);

\end{tikzpicture}
 \caption{The Behler-Parinello like \gls{ann} scheme used to fit total energies and local forces.  A rotation invariant MILAD vector, $\vec{\Phi}_i$, is created for each atomic environment by including neighbours up to the cutoff radius.  Each $\vec{\Phi}_i$ is passed to a feed-forward neural network which outputs the corresponding contribution to the total energy $\epsilon_i$.  Finally, the total energy, $E$, is given by the sum of local contributions.}
 \label{fig:energy-prediction}
\end{figure*}

To give an indication of the performance of MILAD for the prediction of properties we use a Behler-Parinello \citep{Behler2007} like \gls{ann} from the AMP library \citep{Khorshidi2016} for the prediction of total energies and forces.
As shown in \cref{fig:energy-prediction} each atom-centred environment is mapped onto a fingerprint vector, $\vec{\Phi}_i$ that is the input to an \gls{ann} consisting of a number of all-to-all connected linear layers.
The local atomic contributions to the energy are then summed to give the total energy.
We find the hyperbolic tangent activation function to offer a good combination of training speed and accuracy.
The linear layers may be configured with or without a bias.
Our testing suggests that it is preferable to have no bias but rather to rescale the model inputs and outputs to lie in the fixed range (-1, 1) based on the range of values found in the training set.
This also makes it easy to detect when extrapolation is taking place, as in such a case the input or output will fall outside of this range.

Forces are calculated via the chain rule using derivatives of the energy with respect to neural network weights and of the invariants with respect to atomic positions.
To achieve smoothness of the fingerprint vector with respect to atoms entering and leaving the cutoff sphere, the following cosine cutoff function proposed by \citet{Behler2011} is used
\begin{equation}
f(r_{ij})= \begin{cases} 
      \left[ \cos(\frac{\pi r_{ij}}{r_\text{cut}}) + 1 \right] / 2 & r_{ij} \leq r_\text{cut} \\
      0 & r_{ij} > r_\text{cut}
      \end{cases}\text{.}
\end{equation}
The value of this function is used as a prefactor to scale atomic feature weights.

\subsubsection*{Training}

The following loss function is used to train the neural network,
\begin{equation}
 L = \frac{1}{2} \sum_{k=1}^M \frac{1}{N_k} \left[ (E_k - E'_k)^2 + \frac{\alpha}{3} \sum_{i = 1}^{N_k} \sum_{j = 1}^3 (\vec{F}_{i,j} - \vec{F'}_{i,j})^2 \right]\text{,}
 \label{eq:loss}
\end{equation}
where $k$ labels each of the $M$ systems in the training set, $N_k$ is the number of atoms in the $k^\text{th}$ system and primed quantities represent the known training values.
During each training step predictions are made from which the loss function is calculated.
Gradients with respect to the loss, obtained using back-propagation, are then used by the optimiser to update the network weights in an attempt to reduce the value of $L$.
We use the BFGS optimiser throughout.

Given that \glspl{ann} are universal approximators, care must be taken to avoid over-fitting.
Indeed, even random inputs can be successfully \citep{Zhang2016b} learned thus rendering any conclusions based on a training procedure where all of the data is used meaningless.
To avoid this we keep back some of the data as an unseen test set and apply \textit{early stopping}.
This involves calculating the loss for both training and test sets and terminating training when the loss of the test set starts to increase.
In general, we find this situation is much less likely to occur when training with both forces and energies.

\subsubsection*{The experiment}

We use data from two recent studies \cite{Zuo2020,Zeni2021} comparing various descriptors and regression schemes.
The studies compared GAP \cite{Bartok2010,Bartok2015}, \gls{mtp} \cite{Shapeev2016}, a neural network using Behler-Parinello symmetry functions as the descriptor (labelled NNP) \cite{Behler2007,Behler2011}, ACE \cite{Drautz2019}, SNAP \cite{Thompson2015a} and qSNAP \cite{Wood2018}.
The latter two both use hyperspherical harmonics as the basis where the radial component is mapped onto the surface of a four-dimensional sphere.
SNAP and qSNAP both use bispectrum ($\nu = 3$) invariants, the former expressing the energy as a linear expansion while the latter includes quadratic terms.

The data set consists of elemental Ni, Cu, Li, Mo, Si and Ge in a variety of configurations including the ground state crystal structures, strained structures, slabs and \textit{ab initio} molecular dynamics snapshots.
The reference energies and forces were calculated using the VASP density functional theory package and the Perdew-Burke-Ernzerhof functional \cite{Perdew1996}.
For each elemental system there are between 217 and 294 total configurations organised into a 9:1 split of training and test points.

\begin{table}[ht]
\centering
\newcolumntype{M}[1]{>{\centering\arraybackslash\hspace{0pt}}m{#1}}
\begin{tabular}{l M{1.3cm} M{0.8cm} M{0.8cm} M{0.9cm} M{1.6cm} | M{1.6cm}}
 \hline\hline
  & Training point & Test points &  $r_\text{cut}$ (\AA) & $\alpha$ \cref{eq:loss} & Network layers & NNP network layers\\
 \hline
 Ni & 263 & 31 & 3.8 & 0.04 & 32-32 & 24-24 \\
 Cu & 262 & 31 & 3.9 & 0.04 & 24-24 & 8-8 \\
 Li & 241 & 29 & 4.8 & 0.04 & 32-32 & 24-24 \\
 Mo & 194 & 23 & 5.0 & 0.04 & 32-32 & 16-16 \\
 Si & 214 & 25 & 4.7 & 0.04 & 32-32 & 24-24 \\
 Ge & 228 & 25 & 5.1 & 0.05 & 32-32 & 24-24 \\
 \hline\hline
\end{tabular}
\caption{Model parameters for each of the systems studied.  The number of neural network layers for the NNP models from \cite{Zuo2020} are also tabulated for comparison.}
\label{tab:model_parameters}
\end{table}

\Cref{tab:model_parameters} shows the settings used to configure the MILAD descriptor and the neural network for each system.
Limited hyper-parameter tuning was performed to determine the number of neural-network layers to use for each system starting with the same values as reported for the NNP model.
In all cases we found that increasing the number of layers resulted in better accuracy, likely due to the fact that the MILAD fingerprint has more components then the number symmetry functions used by the NNP.

\begin{table*}[ht]
\newcolumntype{R}[1]{>{\raggedleft\arraybackslash\hspace{0pt}}m{#1}}

\begin{tabular}{l R{2.8em} R{2.8em} R{2.8em} R{2.8em} R{2.8em} R{2.8em} p{2.8em} R{2.8em} R{2.8em} R{2.8em} R{2.8em} R{2.8em} R{2.8em}}
\hline\hline
& \multicolumn{6}{c}{Energy RMSD (meV/atom)} & & \multicolumn{6}{c}{Force RMSD (eV/\AA)}\\

& Ni & Cu & Li & Mo & Si & Ge & & Ni & Cu & Li & Mo & Si & Ge \\ 
\hline
GAP & 0.62 & 0.56 & 0.63 & 3.55 & 4.18 & 4.47 & & 0.04 & 0.02 & 0.01 & 0.16 & 0.12 & 0.08 \\ 
MTP & 0.74 & 0.52 & 0.66 & 3.89 & 3.02 & 3.68 & & 0.03 & 0.01 & 0.01 & 0.15 & 0.09 & 0.07 \\ 
NNP & 2.25 & 1.68 & 0.98 & 5.67 & 9.95 & 10.95 & & 0.07 & 0.06 & 0.06 & 0.20 & 0.17 & 0.12 \\ 
SNAP & 1.17 & 0.87 & 1.13 & 9.06 & 8.06 & 10.96 & & 0.08 & 0.08 & 0.04 & 0.37 & 0.34 & 0.29 \\ 
qSNAP & 1.04 & 1.16 & 0.85 & 3.96 & 6.28 & 10.55 & & 0.07 & 0.05 & 0.04 & 0.33 & 0.29 & 0.20 \\ 
3-body ACE \cite{Zeni2021} & 1.74 & 1.19 & 1.23 & 4.00 & 5.16 & 11.62 & & 0.03 & 0.02 & 0.01 & 0.16 & 0.13 & 0.09 \\ 
MILAD & 1.39 & 0.96 & 0.64 & 5.79 & 5.65 & 5.47 & & 0.08 & 0.07 & 0.02 & 0.36 & 0.19 & 0.14 \\ 
\hline
\multicolumn{14}{l}{Empirical potentials}\\
EAM & 8.51 & 7.46 & 368.64 & 67.98 & - & - & & 0.11 & 0.12 & 0.14 & 0.52 & - & - \\
MEAM & 23.04 & 10.49 & - & 36.42 & 111.67 & - & & 0.33 & 0.24 & - & 0.22 & 0.40 & - \\
Tersoff & - & - & - & - & 202.37 & 550.72 & & - & - & - & - & 0.74 & 1.36 \\
\hline\hline
\end{tabular}
\caption{Root-mean-square errors in predicted energies and forces.  Data for the 3-body ACE descriptor are taken from \citet{Zeni2021} while the rest come from \citet{Zuo2020}.}
\label{tab:nn_results}
\end{table*}

\Cref{tab:nn_results} shows the results re-tabulated from \cite{Zuo2020,Zeni2021} along with those obtained using MILAD.
In most cases, the energy RMSDs achieved with MILAD on the test set lie in the range of ACE, NNP, SNAP and qSNAP but, in most cases, higher than GAP and \gls{mtp}.
It is unclear at this stage whether this is due to the difference in regression scheme (GAP uses kernel regression while \gls{mtp} uses linear regression) or the descriptors themselves.
For example, \cref{eq:zernike_from_geometric} allows one to transform from the geometric moments used by \gls{mtp} to Zernike moments used by MILAD suggesting that, at the same correlation order, they should share similar information content.
However, there remain non-trivial differences in the way the radial basis is constructed.
Specifically, \gls{mtp} typically uses a data-driven approach with the basis being tailored to a particular data set rather than being universal.
On the other hand, the training set size is relatively small for neural network regression which tends to be better suited to problems with large amounts of training data.
Comparing to NNP, MILAD achieves lower RMSDs for most of the systems at the cost of requiring layers with more artificial neurons owing to the greater number of fingerprint components.

Looking to the forces the results are similar with MILAD RMSDs generally falling within in range achieved by NNP, SNAP and qSNAP and higher than those achieved by \gls{mtp}, GAP and ACE.
Our testing shows that this trend persists even when increasing the value of $\alpha$ to try and bias the optimiser to converge the forces preferentially over the energies.
It is possible that by performing a grid search over the hyperparameters, as done by \citet{Zuo2020}, that the forces could be converged further still.
An in depth investigation into this and other factors that affect the performance of MILAD for predicting properties will form part of future work.

\section{Conclusion}

In this work we have adapted a method for constructing complete, rotationally invariant, descriptions of finite energy functions from the image analysis community and applied it to the problem of describing atomic environments.
The resulting invariants are algebraically complete, consisting of exactly three fewer equations than the number of moments due to the missing orientation degrees of freedom.
This compactness makes MILAD fingerprints particularly well suited for use as inputs to neural networks where there is a significant training benefit to having a low-dimensional feature space.

The ability to invert MILAD fingerprints to recover atomic environments is particularly appealing and raises the possibility of building generative models that are not based on discontinuous description such as pixels or voxels \citep{Jorgensen2019a}, or linear representations such as SMILES strings \citep{Nesterov2020}.
Furthermore, many generative models depend on the use of an autoencoder that must be trained to find a latent space representation, and this would typically need to be re-trained to be used for each new system.
With our descriptor it is possible to encode a variable number of atoms and atomic species in a fixed length feature vector which is smooth with respect to its inputs.
This enables alchemical models to be built that can be trivially extended to support greater numbers of species, the primary limitation being the effective resolution of the description which can be tuned by increasing the maximum order, $n_\text{max}$.
As the latent space formed by MILAD fingerprints can encode any $f \in L^2$, it remains overcomplete with respect to those $f$ that correspond to sums of atomic feature functions.
As a next step towards building a generative model, work is currently underway to further compress this latent space to find lower dimensional manifolds that correspond to valid atomic configurations.
To be sure, this is an ambitious goal that raises many unsolved issues, it is nevertheless an avenue worth pursuing as it would eliminate the need for data augmentation and ensure that rotational symmetry is respected exactly.

The current set of invariants are limited to a maximum order of $n_\text{max} = 7$ (and therefore $l_\text{max} = 7$), which effectively limits their discriminative power.
This is known to be particularly problematic for highly symmetric environments which require a high angular frequency to be described correctly (in such environments, low order moments are often zero) \cite{Suk2014}.
The limitation of $n_\text{max} = 7$ comes form numerical issues in correctly identifying independent invariants, as standard singular value decomposition methods operate on floating point numbers where a threshold for comparing numbers must be carefully chosen.
This inevitably leads to errors as the number of invariants to be reduced increases.
To overcome this, the method is currently being extended to higher order by using an exact arithmetic library and applying algorithms from the works of \citet{Bachmayr2019} and \citet{Nigam2020}.

In summary, we have shown that MILAD fingerprints can be used for both high-fidelity reconstruction of atomic environments without the need for training and for accurately predicting the properties of atomic systems.
This hints at a novel route to building generative models using existing machine learning tools with a view to enabling general-purpose inverse design of materials and molecules.

\section{Acknowledgements}

I would like to thank Tom\'{a}\v{s} Suk for an enlightening exchange about image moments, Nongnuch Artrith and Andrew Peterson for discussions on neural network based empirical potentials, Peter Bj\o{}rn J\o{}rgensen for discussions about learning architectures and rotation invariants, Alexander Sougaard Tygesen for discussions about recovering atoms from moments and Tess Smidt and Thomas Hardin for discussions on inverting invariants, particularly for suggesting recovering moments from invariants directly.

\section{Data availability}

The codebase for calculating MILAD fingerprints, decoding them back into structures, training neural networks and other related operations can be found at \footnote{\url{https://github.com/muhrin/milad}}.
The notebooks that can be used to reproduce the experiments and most of the images found in this work are located at \footnote{\url{https://github.com/muhrin/milad-paper-2021}}.

The QM9 data set used for reconstruction experiments can be found at \citet{Ramakrishnan2014}.
The dataset for neural network potential fitting can be found at \footnote{\url{https://github.com/materialsvirtuallab/mlearn}}.
An additional neural network fitting experiment found in the Supplemental Material \cite{SI} uses data from \citet{Dragoni2017}.

\nocite{Suk2011}
\nocite{Flusser2016}
\nocite{Kromann2021}
\nocite{Landrum2021}
\nocite{Geiger2021}
\nocite{Lam2015}
\nocite{Meurer2017}
\nocite{Paszke2019}
\nocite{Zuo2020}
\nocite{Zeni2021}
\nocite{Zuo2020}
\nocite{Dragoni2017}
\nocite{Dragoni2018}
\nocite{Kingma2014}
\nocite{Bartok2010}
\nocite{Shapeev2016}

\bibliography{library}

\end{document}


\title{Supplementary information to ``Through the eyes of a descriptor: Constructing complete, invertible, descriptions of atomic environments.''}
\author{Martin Uhrin}
\email{martin.uhrin.10@ucl.ac.uk}
\affiliation{Department of Energy Conversion and Storage, Technical University of Denmark, Kgs. Lyngby DK-2800, Denmark}
\date{\today}

\maketitle

\section{Definitions}

The \textit{associated Legendre functions} are taken to be
\begin{equation}
 P_l^m = (-1)^m (1 - x^2)^{(m / 2)} \frac{d^m}{dx^m}P_l(x)
\end{equation}
where
\begin{equation}
 P_l(x) = \frac{1}{2^l l!}\frac{d^l}{dx^l} (x^2 - 1)^l.
\end{equation}

For transforming between spherical and Cartesian coordinates we use
\begin{eqnarray}
 x =& r \sin(\theta) \cos(\varphi)\text{,} \hspace{4em} & r = \sqrt{x^2 + y^2 + z^2}\text{,} \\
 y =& r \sin(\theta) \sin(\varphi)\text{,} \hspace{4em} & \theta = \arccos(z / r)\text{,} \\
 z =& r \cos(\theta)\text{,} \hspace{4em} & \varphi = \arctan(y / x)\text{.}
\end{eqnarray}

\section{Invariants}

Our code currently uses two sets of rotation invariants.
There is a set of 1185 independent invariants of geometric moments up to 16$^\text{th}$ order based on those of \citet{Suk2011} and 117 invariants of spherical harmonic based moments derived from those of \citet{Flusser2016}, augmented with the following three invariants to reintroduce zero and first order moments:
\begin{align*}
 \Phi_0 = & \Omega_{00}^0\\
 \Phi_1 = & \frac{1}{\sqrt{3}} \left( 2 {c}_{11}^{-1} {c}_{11}^1 - {{c}_{11}^0}^{2} \right)\\
 \Phi_5 = & \frac{\sqrt{5} ({\Omega}_{11}^{-1})^{2} {\Omega}_{22}^{2}}{5} - \frac{\sqrt{10} {\Omega}_{11}^{-1} {\Omega}_{11}^{0} {\Omega}_{22}^{1}}{5} + \frac{\sqrt{30} {\Omega}_{11}^{-1} {\Omega}_{11}^{1} {\Omega}_{22}^{0}}{15}  \\
       & + \frac{\sqrt{30} ({\Omega}_{11}^{0})^{2} {\Omega}_{22}^{0}}{15} - \frac{\sqrt{10} {\Omega}_{11}^{0} {\Omega}_{11}^{1} {\Omega}_{22}^{-1}}{5} + \frac{\sqrt{5} ({\Omega}_{11}^{1})^{2} {\Omega}_{22}^{-2}}{5}\\
 \Phi_6 = & c_1(1, 1)_2 c_2(2, 2)_2 \text{.}
\end{align*}
These account for configurations where the centre of mass is not at the origin \footnote{In the image analysis community it is typical to use central moments to achieve translation invariance in which case all first order moments (and products thereof) are zero, however atom centred environments are not necessarily at the centre of mass.)}.
The full set are sufficient to recover all the corresponding moments up to $7^\text{th}$ order modulo global orientation.

\section{Validation of MILAD fingerprints as a similarity measure}

In order to verify that the RMSD between original and reconstructed fingerprints is a suitable measure of the accuracy of a reconstruction we compare our results against an algorithm that calculates the similarity of two molecules directly based on the atomic coordinates alone.
This is done by finding the rotation and assignment of atoms labels between the two structures that minimises the root-mean-square deviation of the relative atomic positions.
We use a variant of the Hungarian algorithm as implemented in \cite{Kromann2021} to find an initial atom label assignment and then use the RDKit \cite{Landrum2021} library to find the best RMSD, typically using 80,000 attempts.
Ideally, this would form the sole basis our similarity comparison, however, the combinatorial complexity of the algorithm makes it prohibitively expensive beyond a certain number of atoms.
To determine at which point the algorithm fails to find the correct assignment we perform an experiment where we take the original molecule and apply a random rotation and label permutation.
We then calculate the RMSD of the transformed version versus the original.
This is repeated five times for each molecule.
In addition, we calculate the RMSD between each pair of \textit{different} molecules of the same size from the test set containing no species to have a reference value for dissimilar molecules.

\begin{figure*}[!ht]
 \centering
 \includegraphics[width=0.98\textwidth]{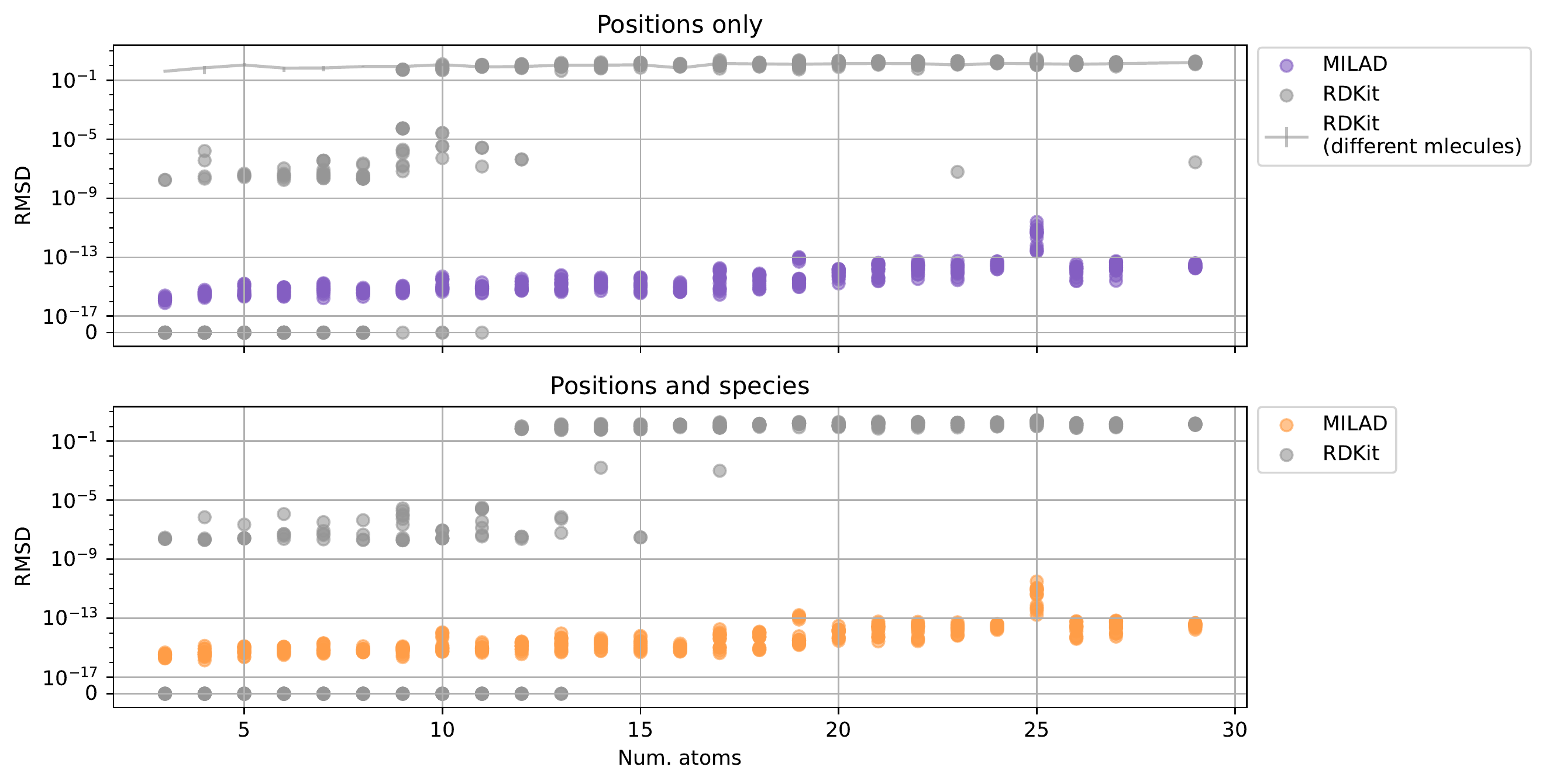}
 \caption{Results showing the RMSD values as given by RDKit when comparing structures directly.  The RMSDs using MILAD fingerprints are also plotted for comparison.}
 \label{fig:rmsd_values}
\end{figure*}

\Cref{fig:rmsd_values} shows that when using positions alone the algorithm can correctly identify transformed molecules up to 8 atoms in size.
Dissimilar molecules have an minimum RMSD of 0.254 \AA.
When species information is included, the number of combinations to check is greatly reduced and the algorithm can correctly identify molecules containing up $N_A = 11$ atoms.

\begin{figure*}[!ht]
 \centering
 \includegraphics[width=0.98\textwidth]{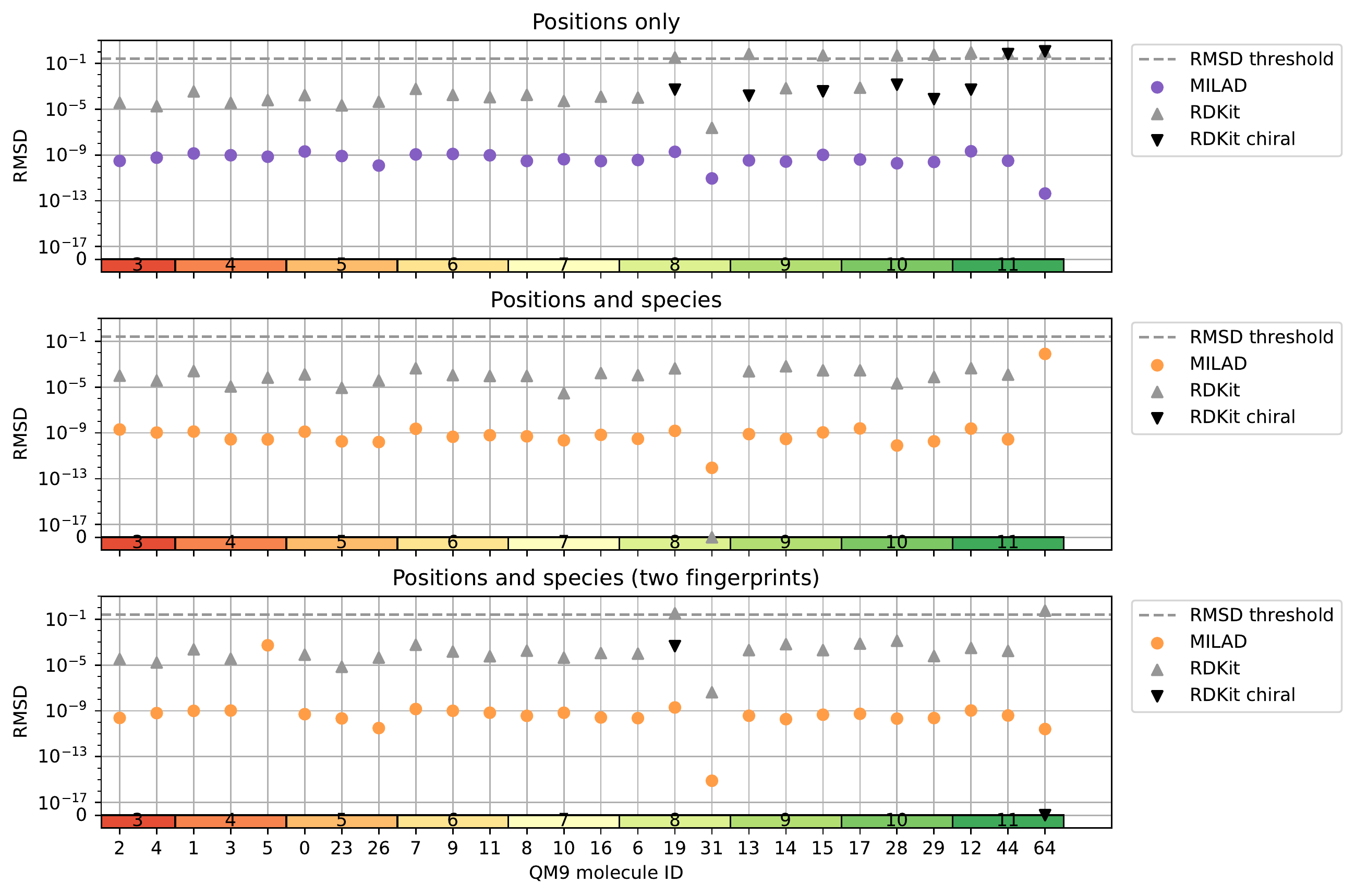}
 \caption{Results comparing using MILAD fingerprints and RDKit RMSDs for judging structural similarity.  Results that fall above the RDKit similarity threshold are re-calculated using the chiral version of the molecule.  The bars at the bottom of each plot show the number of atoms in each molecule.}
 \label{fig:rmsd_comparison}
\end{figure*}

Next, we take each molecule from the reconstruction experiments with up to 11 atoms and use RDKit to calculate the RMSD versus the original.
The results are shown in \cref{fig:rmsd_comparison}.
For positions only, we see that both RDKit and MILAD fingerprints agree on successful reconstruction up to $N_A = 8$, above which there are several which fall near or above the RDKit threshold.
These are, in fact, chiral versions of the original molecules.
When we perform an inversion about origin and recalculate the RDKit RMSD it inevitably falls below the threshold again (with the exception of $N_A = 11$, however we know from \cref{fig:rmsd_values} that is it not possible to reliably calculate RMSDs for molecules with $N_A > 8$ when all species are the same).
This highlights an important limitation of the MILAD.
By keeping only the scalars corresponding to the $\textbf{D}^0$ representation the descriptor is invariant to reflections and is incapable of distinguishing chiral molecules.
To overcome this it would be necessary to keep track of parity as is commonly done in equivariant neural networks (see e.g. \cite{Geiger2021}).

Both sets of results with species included show a consistent agreement between MILAD fingerprints and RDKit RMSDs when judging structural similarity up to $N_A = 11$ once chiral versions are included.

\section{Implementation}

The entire codebase created for this work is open source and available on github.
It is implemented in Python which allows for fast development and prototyping as well as easy integration with existing Python codebases.
The Numba \citep{Lam2015} library has been used for performance-critical sections which gives good performance thanks to just-in-time compilation.
This was found to be competitive with, if slightly slower then, a similar Julia implementation.
Much of the codebase has unit tests, and, where relevant, the Sympy \citep{Meurer2017} symbolic mathematics library has been use the ensure the correctness of algorithms.
The neural network model is implemented using PyTorch \citep{Paszke2019} and can therefore run efficiently on CPUs, GPUs or TPUs.

\section{Reconstruction experiment}

The IDs of the subset of molecules used form the QM9 data set are:

\setlength\extrarowheight{-5pt}
\begin{tabular}{cccc}
Num. atoms & \multicolumn{3}{c}{IDs} \\
\hline\hline
3: & 2 & 4 \\
4: & 1 & 3 & 5 \\
5: & 0 & 23 & 26 \\
6: & 7 & 9 & 11 \\
7: & 8 & 10 & 16 \\
8: & 6 & 19 & 31 \\
9: & 13 & 14 & 15 \\
10: & 17 & 28 & 29 \\
11: & 12 & 44 & 64 \\
12: & 21 & 39 & 40 \\
13: & 62 & 67 & 71 \\
14: & 20 & 38 & 101 \\
15: & 54 & 80 & 83 \\
16: & 218 & 223 & 226 \\
17: & 53 & 82 & 132 \\
18: & 225 & 228 & 229 \\
19: & 1081 & 1083 & 1087 \\
20: & 227 & 273 & 290 \\
21: & 1091 & 1094 & 1095 \\
22: & 5796 & 5809 & 5812 \\
23: & 1093 & 1103 & 1129 \\
24: & 5806 & 5807 & 5808 \\
25: & 36927 & 36945 & 36959 \\
26: & 5805 & 5810 & 5850 \\
27: & 42138 & 57349 & 57419 \\
29: & 57517 & 58098 & 58182 \\
\end{tabular}

\section{Ni, Cu, Li, Mo Si and Ge neural network potentials}

Here we give the full training and test results for the neural network comparison to the benchmarks of \citet{Zuo2020} and \citet{Zeni2021}.

\begin{table}
\scriptsize
\newcolumntype{R}[1]{>{\raggedleft\arraybackslash\hspace{0pt}}m{#1}}

\begin{tabular}{l p{4.5em} R{2.8em} R{2.8em} R{2.8em} R{2.8em} R{2.8em} R{2.8em} p{2.8em} R{2.8em} R{2.8em} R{2.8em} R{2.8em} R{2.8em} R{2.8em}}
& & \multicolumn{6}{c}{Energy RMSD (meV/atom)} & & \multicolumn{6}{c}{Force RMSD (eV/\AA)}\\

& & Ni & Cu & Li & Mo & Si & Ge & & Ni & Cu & Li & Mo & Si & Ge \\ 
\hline
\multirow{2}{4em}{GAP} & Training & 0.62 & 0.60 & 0.75 & 3.01 & 4.49 & 4.02 & & 0.03 & 0.01 & 0.01 & 0.15 & 0.10 & 0.07 \\ 
 & Test & 0.62 & 0.56 & 0.63 & 3.55 & 4.18 & 4.47 & & 0.04 & 0.02 & 0.01 & 0.16 & 0.12 & 0.08 \\ 
\hline 
\multirow{2}{4em}{MTP} & Training & 0.66 & 0.52 & 0.73 & 3.53 & 3.24 & 4.46 & & 0.02 & 0.01 & 0.01 & 0.14 & 0.07 & 0.06 \\ 
 & Test & 0.74 & 0.52 & 0.66 & 3.89 & 3.02 & 3.68 & & 0.03 & 0.01 & 0.01 & 0.15 & 0.09 & 0.07 \\ 
 \hline
\multirow{2}{4em}{NNP} & Training & 2.38 & 2.13 & 1.24 & 6.06 & 10.84 & 11.27 & & 0.06 & 0.05 & 0.06 & 0.20 & 0.17 & 0.12 \\ 
 & Test & 2.25 & 1.68 & 0.98 & 5.67 & 9.95 & 10.95 & & 0.07 & 0.06 & 0.06 & 0.20 & 0.17 & 0.12 \\ 
\hline
\multirow{2}{4em}{SNAP} & Training & 1.59 & 1.04 & 1.39 & 8.45 & 8.10 & 11.82 & & 0.07 & 0.07 & 0.04 & 0.37 & 0.31 & 0.29 \\ 
 & Test & 1.17 & 0.87 & 1.13 & 9.06 & 8.06 & 10.96 & & 0.08 & 0.08 & 0.04 & 0.37 & 0.34 & 0.29 \\ 
\hline
\multirow{2}{4em}{qSNAP} & Training & 0.82 & 1.41 & 1.10 & 4.68 & 6.99 & 10.44 & & 0.06 & 0.04 & 0.04 & 0.32 & 0.26 & 0.19 \\ 
 & Test & 1.04 & 1.16 & 0.85 & 3.96 & 6.28 & 10.55 & & 0.07 & 0.05 & 0.04 & 0.33 & 0.29 & 0.20 \\ 
 \hline
 3-body ACE \cite{Zeni2021} & Test & 1.74 & 1.19 & 1.23 & 4.00 & 5.16 & 11.62 & & 0.03 & 0.02 & 0.01 & 0.16 & 0.13 & 0.09 \\ 
\hline 
\multirow{2}{4em}{MILAD} & Training & 0.89 & 0.68 & 0.44 & 4.96 & 2.44 & 3.42 & & 0.05 & 0.04 & 0.01 & 0.22 & 0.11 & 0.11 \\  
 & Test & 1.39 & 0.96 & 0.64 & 5.79 & 5.65 & 5.47 & & 0.08 & 0.07 & 0.02 & 0.36 & 0.19 & 0.14 \\ 
 \hline
 \multicolumn{15}{l}{Empirical potentials}\\
 \multirow{2}{4em}{EAM} & Training & 8.02 & 7.29 & 339.46 & 76.06 & - & - & & 0.10 & 0.11 & 0.13 & 0.54 & - & - \\
  & Test & 8.51 & 7.46 & 368.64 & 67.98 & - & - & & 0.11 & 0.12 & 0.14 & 0.52 & - & - \\
  \hline  
 \multirow{2}{4em}{MEAM} & Training & 20.37 & 9.48 & - & 32.62 & 104.58 & - & & 0.29 & 0.20 & - & 0.23 & 0.35 & - \\
  & Test & 23.04 & 10.49 & - & 36.42 & 111.67 & - & & 0.33 & 0.24 & - & 0.22 & 0.40 & - \\
  \hline
 \multirow{2}{4em}{Tersoff} & Training & - & - & - & - & 184.83 & 561.37 & & - & - & - & - & 0.67 & 1.38 \\
  & Test & - & - & - & - & 202.37 & 550.72 & & - & - & - & - & 0.74 & 1.36 \\
\end{tabular}

\caption{Root-mean-square errors in predicted energies and forces.  Data for the 3-body ACE descriptor are taken from \citet{Zeni2021} while the rest come from \citet{Zuo2020}.}
\label{tab:nn_results}
\end{table}

\section{Iron NN-potential}

This experiment appeared in earlier versions of the manuscript but has been moved here as a new set of experiments that give a broader comparison of MILAD's performance now appears in the main manuscript.

We use training data from \citet{Dragoni2017,Dragoni2018} which includes DFT forces and energies for iron in a variety of atomic configurations.
For this experiment we use database 1 which includes distorted primitive (single atom) unit cells and database 2 which contains snapshots from NVT molecular dynamics runs of $3 \times 3 \times 3$ (54 atoms) and $4 \times 4 \times 4$ (128 atoms) conventional cubic supercells.
We take randomly shuffled subsets of these databases and keep 90\% for training and 10\% for testing.
The training is performed in batches where a subset of the training data is passed through the network after which the weights are updated by the optimiser, in each training epoch all training batches are used.
We find this scheme produces results with a lower final loss and postpones the onset of overfitting, albeit at a higher computational cost.
Pertinent parameters relating to the experiment are tabulated below.

\begin{center}
{\small
\setlength\extrarowheight{-10pt}
\begin{tabular}{l r}
 Total systems & 400 (200 DB1, 200 DB2)\\
 Training set & 360 \\
 Test set & 40 \\
 Invariants & Zernike (to $n_\text{max} = 7$), 117 total \\
 Feature function & delta \\
 Hidden layers & 32-16-8 \\
 Cutoff & 3.5 \AA \\
 Activation fn. & Tanh\\
 Batch size & 128-512 \\
 Optimiser & Adam \citep{Kingma2014} \\
 $\alpha$ & 0.0005 \\
 
\end{tabular}
}
\end{center}

\begin{figure*}[!ht]
     \centering
     \subfloat[Energy differences versus DFT data.]{
         \centering
         \includegraphics[width=0.48\textwidth]{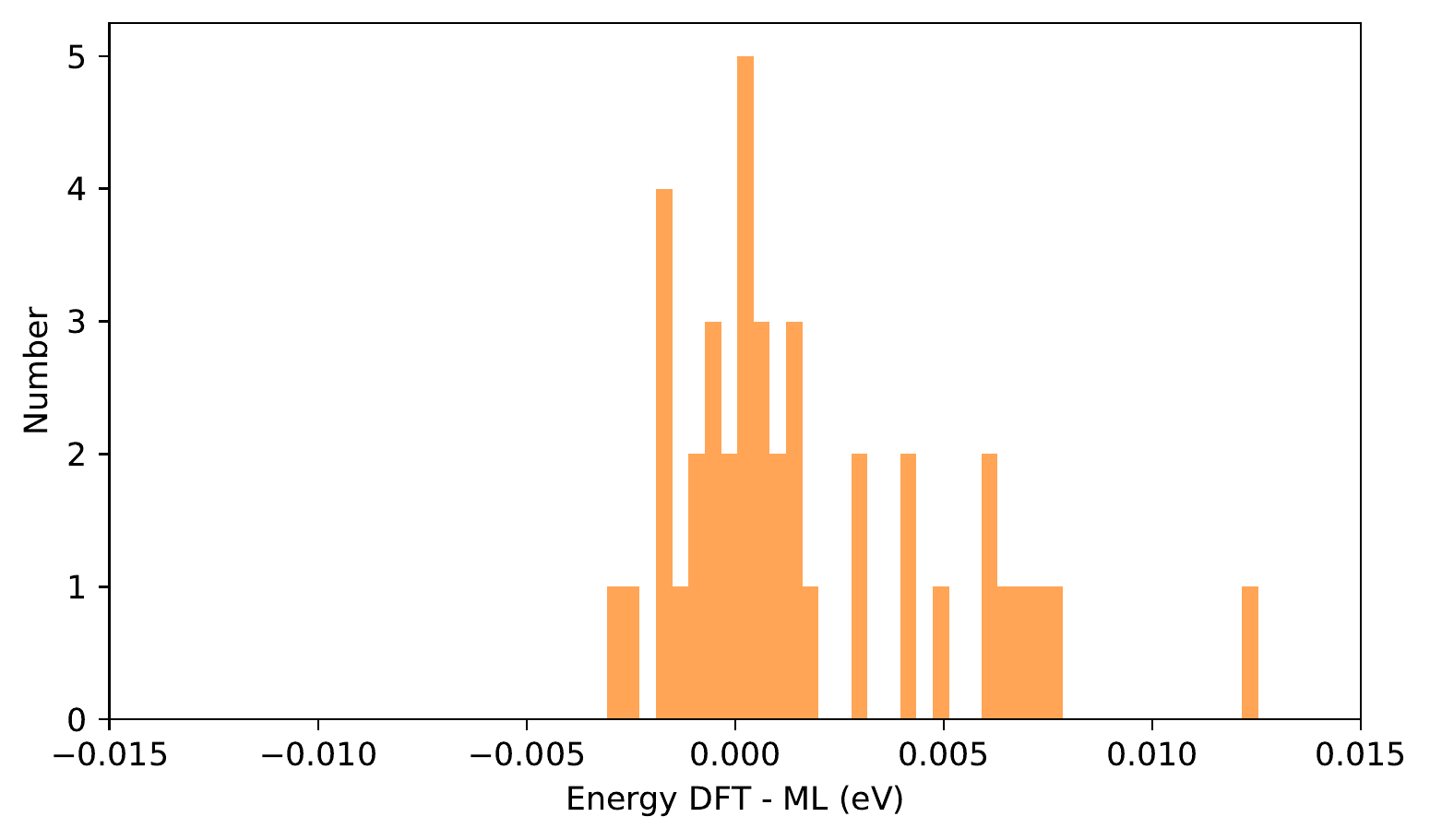}
         }
     \subfloat[Force component differences versus DFT data]{
         \includegraphics[width=0.48\textwidth]{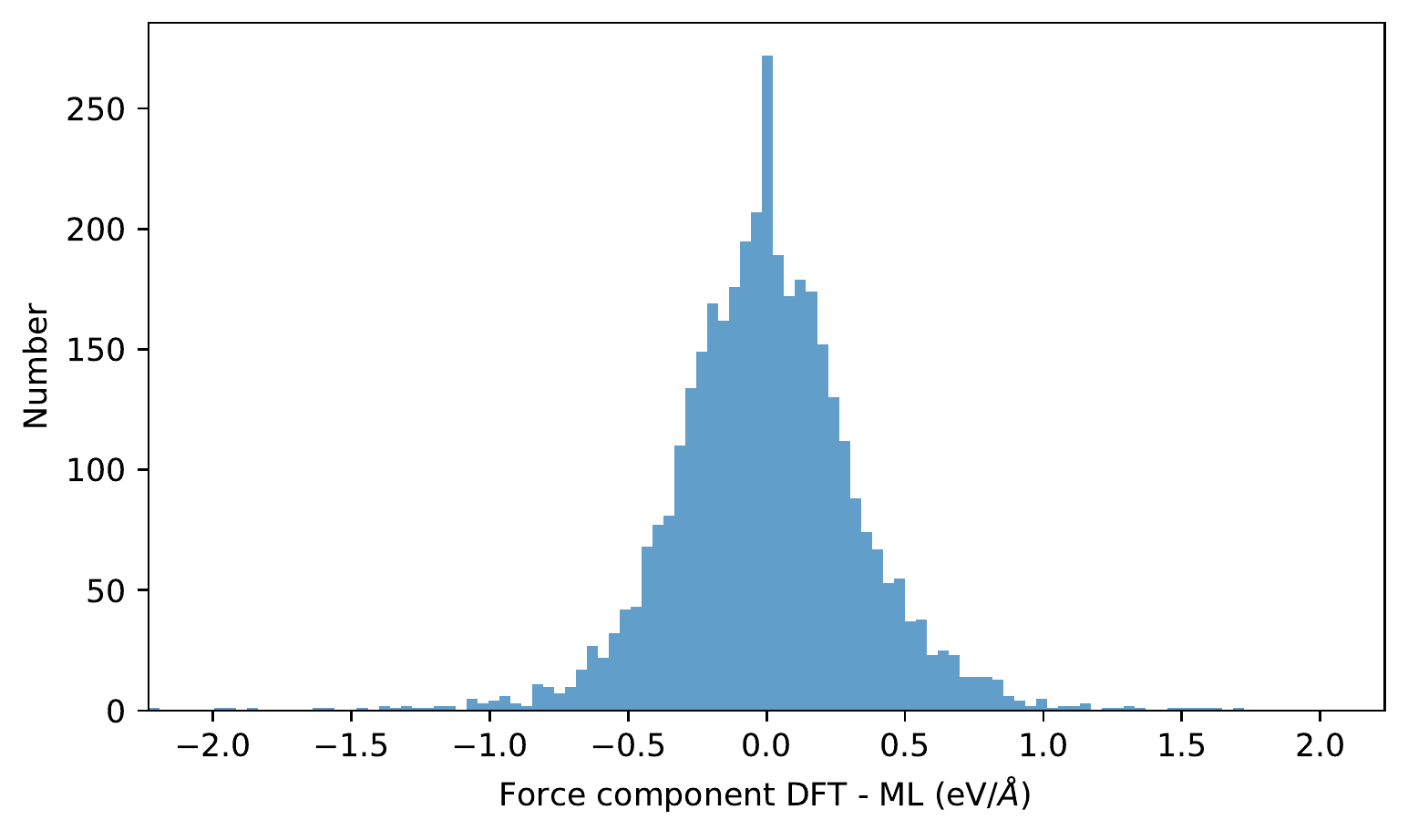}
     }
  \caption{Plots of energy and force differences for the test set as predicted by the neural network model versus ground truth DFT data.}
  \label{fig:nn-results}
\end{figure*}

\Cref{fig:nn-results} shows results of force and energy deviations from the ground truth on the unseen test set.
Overall root-mean-square deviations of 3.703 meV/atom and 0.197 eV/\AA{} were achieved for energies and forces respectively.
Energy deviations are generally clustered to within 3 meV of DFT with some asymmetry and a single outlier around 12 meV while force errors are well distributed around the origin.
Our model shows results similar to those of a recent comparison of machine learning based interatomic potentials \citep{Zuo2020} where the neural network model achieved energy RMSDs of between 0.98 to 10.95 meV/atom and force RMSDs of between 0.06 to 0.2 eV/\AA{} depending on the material.
The best models in that study, Guassian approximation potentials \citep{Bartok2010} and moment tensor potentials \citep{Shapeev2016} perform better still with energy RMSDs of between 0.52 and 4.47 meV/atom, and force RMSDs of between 0.01 and 0.16 eV/\AA{}.
It is likely that by extending our invariants to orders higher than the current limit of $n_\text{max} = 7$ the prediction accuracy of our models will improve further and this will form a part of future work.

\bibliography{library}